\documentclass[smallextended]{svjour3}       % onecolumn (second format)

\usepackage{multirow}
\usepackage{balance}
\usepackage{graphicx}
\usepackage{alltt}
\usepackage{relsize}
\usepackage{booktabs}
\usepackage{array}
\usepackage{amsmath}
\usepackage{verbatim}

\usepackage[tight,footnotesize]{subfigure}

\usepackage{amsfonts}
\usepackage{amssymb}
\usepackage{boxedminipage}
\usepackage{float}
\usepackage{fancyvrb}
\usepackage[dvipsnames]{xcolor}
\usepackage[colorlinks=true]{hyperref}
\hypersetup{
linkcolor=blue
,citecolor=blue
,filecolor=blue
,urlcolor=blue
,menucolor=blue
,runcolor=blue
,linkbordercolor=blue
,citebordercolor=teal
,filebordercolor=blue
,urlbordercolor=blue
,menubordercolor=blue
,runbordercolor=blue
}
\usepackage{balance}
\usepackage{url}
\usepackage{fancybox}%for \hypobox
\usepackage{listings}
\usepackage{array}
\usepackage{lscape}
\usepackage{textcomp}

\usepackage{makecell}
\usepackage{rotating}
\definecolor{mygray}{rgb}{0.7,0.7,0.7}
\newcommand{\heng}[1]{\textcolor{red}{{\it [Heng: #1]}}}
\newcommand{\xingfang}[1]{\textcolor{teal}{{\it [Xingfang says: #1]}}}

\newcommand{\responseminor}[2]{#2}
\newcommand{\pararesponseminor}[1]{#1}

\newcommand{\response}[2]{#2}
\newcommand{\pararesponse}[1]{#1}
\newcommand{\accepted}[1]{}

{\begin{list}{}%
         {\setlength{\leftmargin}{#1}}%
         \item[]%
}
{\end{list}}

{ \medskip
  \noindent
  \let\emph=\textbf
  \begin{boxedminipage}{\columnwidth}\em}%
{ \end{boxedminipage}}
\newdimen\qdx
\newdimen\qda
\newdimen\qdb
\def\rrrr#1#2#3#4{\newdimen\qd\qd=#4 % length of bar for 1.0
\qdx=\qd\multiply\qdx by 5\divide\qdx by 4
\qda=\qd
\qdb=\qd
\multiply\qda by #1\divide\qda by #3\multiply\qdb by #2\divide\qdb by #3\advance\qdb by -\qda
    \leavevmode\hbox to \qdx{\hfil\vbox{%
    \hbox{\vrule\vbox{\hrule\hbox to 1\qd
            {\vrule depth0pt height0.7ex width \qda\color{mygray}%
 \vrule depth0pt height0.7ex width \qdb\hfill}\hrule}\vrule}
    }\hfil}}

  \renewenvironment{thebibliography}[1]{%
    \begin{oldthebibliography}{#1}%
      \setlength{\parskip}{0ex}%
      \setlength{\itemsep}{0ex}%
  }%
  {%
    \end{oldthebibliography}%
  }
\newcommand{\footnoteref}[1]{\textsuperscript{\ref{#1}}}

\usepackage{graphicx}

\usepackage{framed}
\usepackage{lscape}
\usepackage{setspace}
\usepackage{tabularx}
\usepackage{multirow}
\usepackage{stfloats}
\usepackage[normalem]{ulem}

\usepackage{url}
\usepackage[flushleft]{threeparttable}
\usepackage{comment}
\usepackage[round,sort]{natbib}

\usepackage{pgfplots}
\pgfplotsset{compat=1.15}
\usepackage{tikz}
\usetikzlibrary{positioning}
\usepackage{subfigure}

\makeatletter
\renewcommand\hyper@natlinkbreak[2]{#1}
\makeatother
\def\makeheadbox{{%
\hbox to0pt{\vbox{\baselineskip=10dd\hrule\hbox
to\hsize{\vrule\kern3pt\vbox{\kern3pt
\hbox{\bfseries Empirical Software Engineering}
\hbox{This is a pre-copyedit version of this article.}
\hbox{The final version is available online at:~DOI \href{https://doi.org/10.1007/s10664-023-10364-1}{10.1007/s10664-023-10364-1}}
\kern3pt}\hfil\kern3pt\vrule}\hrule}%
\hss}}}

\begin{document}

\title{On the Effectiveness of Log Representation for Log-based Anomaly Detection
}
% \subtitle{Do you have a subtitle?\\ If so, write it here}

\titlerunning{On the Effectiveness of Log Representation}        % if too long for running head

\author{Xingfang Wu         \and
        Heng Li \and
        Foutse Khomh    %etc.
}

% \author{Anonymous submission \heng{include author information}}

\institute{Xingfang Wu, Heng Li, Foutse Khomh \at
              Department of Computer Engineering and Software Engineering \\
              Polytechnique Montreal \\
              Montreal, QC, Canada \\
              \email{\{xingfang.wu, heng.li, foutse.khomh\}@polymtl.ca}
}

\date{Received: date / Accepted: date}

\maketitle

\begin{abstract}
%\secdescription{Leave it to the end.}

Logs are an essential source of information for people to understand the running status of a software system. Due to the evolving modern software architecture and maintenance methods, more research efforts have been devoted to automated log analysis.
%proposing and refining algorithms that can automatically analyze system logs. 
In particular, machine learning (ML) has been widely used in log analysis tasks. In ML-based log analysis tasks, converting textual log data into numerical feature vectors is a critical and indispensable step. 
%Prior work uses different log representation techniques in their ML-based log analysis models.
However, the impact of using different log representation techniques on the performance of the downstream models is not clear, which limits researchers and practitioners' opportunities of choosing the optimal log representation techniques in their automated log analysis workflows.
Therefore, this work investigates and compares the commonly adopted log representation techniques from previous log analysis research. Particularly, we select six log representation techniques and evaluate them with seven ML models and four public log datasets (i.e., HDFS, BGL, Spirit and Thunderbird) in the context of log-based anomaly detection. We also examine the impacts of the log parsing process and the different feature aggregation approaches when they are employed with log representation techniques. From the experiments, we provide some heuristic guidelines for future researchers and developers to follow when designing an automated log analysis workflow. We believe our comprehensive comparison of log representation techniques can help researchers and practitioners better understand the characteristics of different log representation techniques and provide them with guidance for selecting the most suitable ones for their ML-based log analysis workflow.

% \xingfang{https://github.com/mooselab/suppmaterial-LogRepForAnomalyDetection}


%\subclass{MSC code1 \and MSC code2 \and more}
\end{abstract}

\section{Introduction} \label{sec:introduction}

% Follow the flow:
% 1. Importance of logs and log analysis.
% 2. Challenges of log analysis (e.g., big and complex data) and importance of automated approaches (e.g., automated anomaly detection).
% 3. The concept of vectorizing logs and its application in automated log analysis. Give some examples. Highlight its importance. Give a formal definition of log representation.
% 4. States the problem that you want to address: prior studies use a variety of log representation techniques. It is not clear whether and how these representation techniques impact the performance of the downstream tasks.
% 5. Give a summary of your approach and introduce your research questions. For each RQ, briefly summarize the motivation, approach, and results.
% RQ1: 
% RQ2: 
% RQ3:
% 6. Summarize your findings and their implications.
% 7. Introduce the organization of the rest of the paper.
%\accepted{\Foutse{the intro i lacking references! you need to back up some of the statements}}
Logs are textual data generated by logging statements in the source code of software systems. Log data records important runtime information so that software practitioners can use it to understand the running state of a software system or diagnose a system failure. Traditionally, developers and operators manually examine logs or use rule-based approaches to search and analyze log data~\citep{hansen1993automated,prewett2003analyzing,rouillard2004real}, which proves to be very inefficient and error-prone ~\citep{oliner2012advances}. %\accepted{\Foutse{add a reference?}}
Modern software systems are large-scale, especially distributed systems that run on thousands of commodity machines, which usually generate large volumes of logs each day~\citep{oliner2007supercomputers, Schroeder07DiskFailures}.
%making manual checking become impossible. 
Logs are usually semi-structured and exhibit a mixture of formats and vocabularies, making the traditional manual or rule-based approaches tremendously challenging, if not infeasible~\citep{zhu2019tools, dai2020logram}. 
Furthermore, structures and maintenance practices %\accepted{\Foutse{practices?}} 
of modern software systems change rapidly, which poses new challenges %\accepted{\Foutse{elaborate a bit?}} 
for log analysis~\citep{shang2014exploratory,yuan2012characterizing}. 
Automated log processing has drawn many software engineering researchers’ interest in this context.

Prior studies have proposed various approaches that leverage information retrieval, natural language processing, traditional machine learning, and deep learning to support automated log analysis tasks~\citep{he2021survey}. Automated log analysis approaches have been playing an important role in software maintenance and operation efforts (e.g., anomaly detection~\citep{fu2009execution, xu2009detecting, he2016experience, chen2021experience, meng2019loganomaly, le2021log, wang2018anomaly, zhang2019robust, du2017dl, lu2018detecting, nedelkoski2020self}, failure diagnosis~\citep{fu2013contextual, yuan2010sherlog}, performance regression analysis~\citep{nagaraj2012structured, chow2014mystery, liao2020using}). Many of these automated log analysis tasks leverage machine learning (ML) techniques. 
An indispensable step of ML-based log analysis is to transform the textual log data into numerical formats (e.g., feature vectors or digital sequences) that ML models can consume as features. 
We refer to this step as \textbf{log representation}: the process that transforms textual log data into numerical formats to be used as features in ML models. 
%Log representation, an essential step in transforming and vectorizing textual log data into representations (e.g., vectors, digital sequence, etc.) that analytical models can consume, serves as an interface for various downstream tasks~\citep{sadeghi2020log}.

Prior work uses different log representation techniques in their ML-based log analysis tasks (i.e., downstream tasks), including classical techniques (e.g., counting the occurrences of log templates or TF-IDF) and (deep) neural network based techniques (e.g, Word2Vec or FastText).
For example,~\citet{he2016experience} match Message Count Vector representation with a logistic regression model to detect anomalies in log sequences.~\citet{zhang2019robust} leverages pre-trained FastText model to generate log template embeddings to construct their anomaly detection workflow.
%\heng{give two example papers: one for classical message count vector, another for NN-based embedding like word2vec}.
However, no work has focused on evaluating the effectiveness of these representations, thus the impact of using different log representation techniques on the performance of the downstream models is not clear.
Although there are some ablation studies of automated log analysis to evaluate the effectiveness of their adopted representations for log data~\citep{chen2021experience}, researchers can hardly compare the studied log representation techniques with that of other works to know about the impacts that these techniques may have on the performance of downstream tasks.
%, which limits researchers and practitioners' opportunities of choosing the optimal log representation techniques in their automated log analysis workflows. %,  there is a lack of a comprehensive comparison among the different log representation techniques, and  . 
%In this situation, researchers and practitioners can only get a rough impression about various representation techniques from the performances of different log analysis models, which may have biases due to the implementations and erroneous intuitions. 
%Therefore, this work investigates and compares the commonly adopted log representation techniques from previous log analysis research. 
%Prior studies usually use different approaches of log representation 
Therefore, our work aims to provide a comprehensive investigation of log representation techniques with the goal of providing a reference for future research on automated log analysis. 
We select six commonly used log representation techniques and evaluate them with seven
%nine\heng{seven after removing unsupervised ones}
ML models and four public log datasets in the context of log-based anomaly detection task.
%We select common log representation techniques and pre-trained embeddings from related works and evaluate the log representations generated with anomaly detection, 
We select the context of log-based anomaly detection as it is the most widely studied topic of automated log analysis~\citep{fu2009execution, xu2009detecting, he2016experience, chen2021experience, meng2019loganomaly, le2021log, wang2018anomaly, zhang2019robust, du2017dl, lu2018detecting, nedelkoski2020self}. \response{R3.2}{Our findings are likely to be generalizable to other automated log analysis tasks, given the similarity among log representation techniques used in various downstream tasks~\citep{he2021survey}\accepted{\heng{Would be good to have some references here, such as ``A Survey on Automated Log Analysis for Reliability Engineering
'' by He et al.}}. Therefore, the key factors we identified for selecting log representation techniques are expected to hold for other automated log analysis downstream tasks as well.}
%We aim to know about the characteristics and measure the effectiveness of these techniques by their performances on anomaly detection.
We achieve our research objectives by answering the following research questions (RQs): 
%\accepted{\heng{rephrased the RQ titles, keep it consistent in the paper}
%\heng{For each RQ, briefly summarize the motivation, approach, and results}
%\heng{move/merge the following three paragraphs to below each RQ title.}}

\noindent $\bullet$ \textbf{RQ1: How effective are existing log representation techniques for automated log analysis?}

This research question aims at making a fair comparison of the existing common log representation techniques. In this research question, we combine different log representation techniques with different anomaly detection models. By comparing and analyzing the performances across the combinations, we derive some observations for developers and researchers to help better choose log representation techniques when designing automated log analysis frameworks.

\noindent $\bullet$ \textbf{RQ2: How does log parsing influence the effectiveness of log representations in automated log analysis?}

Log parsing is a common pre-processing step before the log representation step. It is not clear how log parsing and log representation together impact the performance of downstream tasks. 
%, may generate inaccurate results. Some automated log analysis frameworks abandon the log parser and take raw logs as inputs. 
Thus, in this RQ, we investigate the potential impacts that log parsing, when used with different log representation techniques, may have on the performance of downstream models. Findings confirm that the log parsing process has non-negligible impacts on the performance of the downstream models.

\noindent $\bullet$ \textbf{RQ3: How do representation aggregation methods influence the effectiveness of log representation in automated log analysis?}

% \xingfang{Minor adjustments to the following paragraph}
Log representation techniques can generate the representation at different levels (e.g., token level or log event level). Sometimes, low-level representations need to be merged into high-level ones according to the need of the follow-up models. In this RQ, we aim to explore the potential influence of different aggregation configurations when used together with different log representation techniques. The findings indicate that the impacts of aggregation configurations may vary according to different factors, and the aggregation configurations may have non-negligible influences on the quality of log representations. 
% Therefore\heng{maybe remove ``Therefore'' as you did not mention not single best solution before this point}, r
Researchers should be careful when doing feature aggregation as there is no single best solution for all log data, representation techniques, and models.

%By answering these research questions, we hope our work can provide a concise reference for related researchers and software practitioners \accepted{\Foutse{practitioners too...}} to select suitable representation techniques for their log analysis tasks according to the characteristics of data and downstream models. We also expect that our work can inspire them \accepted{\Foutse{how? it would be nice if we could report some concrete guidelines? highlight some specific area for improvement?}\xingfang{It is a little bit hard to specify very concrete guidelines according to the experiments we have. But I try to mention some potential aspects.}} to design more effective log representation techniques by considering the special characteristics of log data to decide the way of representing dynamic fields in log data and the granularity of information.
%\sout{To be concrete, we investigate the different combinations of anomaly detection models with our studied anomaly detection models. We analyze the results and propose the directions for future researchers to choose suitable log representation techniques. Further, we find that the log parsing process greatly impacts the log representation. When designing an automated log analysis, researchers should be careful when deciding whether to adopt a log parser. At last, we evaluate different configurations of feature aggregation in log representation and find that aggregation approaches have non-negligible influences on the effectiveness of log representation.} 

Our work makes several important contributions:
\begin{enumerate}
    % \vspace{-2mm}
    \item We provide a comprehensive evaluation of the impact of log representation techniques on log-based anomaly detection task. Our results can be used as a guide for researchers and software practitioners in selecting the most suitable log representations for their anomaly detection frameworks or other log analysis workflows.
    \item We provide an analysis of the impact of log parsing and feature aggregation approaches when they are used together with different log representation techniques. The insights obtained through this analysis can help optimize workflows of log analysis.
    \item We share an implemented pipeline for log-based anomaly detection which supports convenient configurations of log parsing, different log representations, and different aggregation methods.
    % \heng{make sure these are done in the replication package: allows users to conveniently configure the pipeline.}\xingfang{yes, all the steps are included in the package.} 
    Our implementation of the pipeline together the steps to replicate our main results are included in our replication package\footnote{\label{reppack}Scripts and data files used in our research are available online and can be found in our replication package: \newline \url{https://github.com/mooselab/suppmaterial-LogRepForAnomalyDetection}.}. % so that future work can replicate or build on our work.
\end{enumerate}

\noindent \textbf{Organization.} The remainder of this paper is organized as follows: We introduce the background of our work in Section \ref{sec:background}. Section \ref{sec:related} surveys related works. Section \ref{sec:design} describes the design of our experiments, including the selection and overview of studied log representations, the downstream task and the datasets used. The evaluation metrics for the downstream task are also introduced. Section \ref{sec:results} is organized by the research questions we proposed. For each research question, we present the corresponding approach and results. Section \ref{sec:discussions} discusses our findings from the three research questions and summarizes the take-home messages.
Section~\ref{sec:threats} identifies the threats to validity of our findings. At last, we summarize this paper in Section~\ref{sec:conclusions}.

\section{Background} \label{sec:background}

%\secdescription{Introduce the background to help readers understand your study.}

\subsection{Log representations}
%\secdescription{
%- Formal definition of log representation (in an abstract way).
%- Describe the different types of log representations used in your study.}

\begin{figure}[ht]
    \centering
	\includegraphics[scale=0.38]{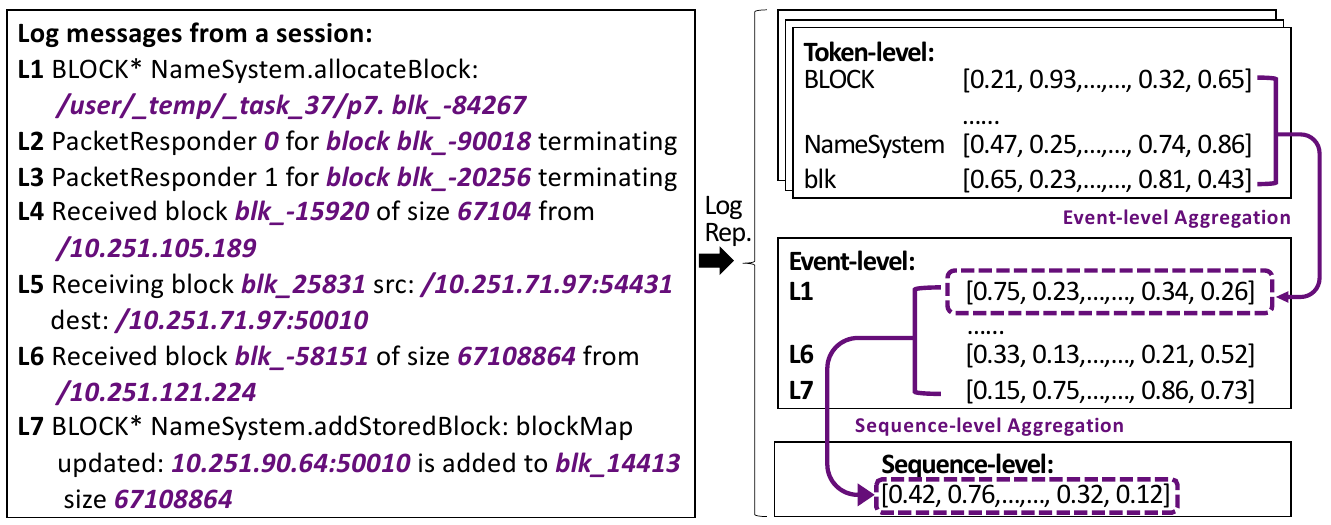}
	\caption[Level]{Different levels of abstraction of log representation.\response{R2.1-Graph updated to show the relations between different levels of representation.}{}\accepted{\heng{if the figure is updated, the title can be colored to indicate the change}}}
	\label{fig:replevel}
\end{figure}

Log representation is a process that converts textual log data into numerical feature vectors. %\heng{would be good to use a small figure to illustrate what is log representation.}\xingfang{add later.} 
Log representation techniques take semi-structured raw log data or parsed log data as input and generate representations at different abstraction levels. Figure \ref{fig:replevel} illustrates an example of different levels of representation for a log session. Different log representation techniques may work on different levels. Aggregation is the process that merges low-level representation into high-level one. Based on these different levels of representations, various follow-up models can be designed to perform downstream tasks according to the needs. Different log analysis tasks may work on different levels of log representation according to their needs of information.

%There are different levels of abstractions of log representation in automated log analysis. Different levels of log abstractions are encoded into feature vectors, whereby a machine learning model can be applied. 

\noindent \textbf{Token-level representation.}  One piece of log message itself is a sequence of tokens. Tokens can be represented as embeddings with pre-trained language models (e.g., word2vec). We call this level of abstraction as \textbf{token-level} representation, which is the lowest level of log representation. \response{R2.1}{Instead of directly feeding the token-level representation to the follow-up models to fulfill downstream tasks, representations of this level are usually aggregated into higher-level\accepted{\heng{higher-level}} ones, where the aggregation techniques are applied. However, there exist log anomaly detection methods that mainly work on token-level representations. For example, Logsy~\citep{nedelkoski2020self} tokenizes the preprocessed log messages and generates embeddings for tokens in log templates. \accepted{\heng{small change here}}Together with the positional encoding of the tokens, these representations are fed into a transformer-based structure. By training the neural network, the token-level representation is updated.} 

\noindent \textbf{Event-level representation.}
A log \textbf{event-level} embedding is a vector representation that encodes a single log message. This level of representation can be merged from token-level embeddings with different aggregation approaches~\citep{meng2019loganomaly}. Besides, some language models can directly generate this level of representation directly~\citep{devlin2018bert,le2021log}. \response{R2.1}{For example, Swisslog~\citep{li2020swisslog} employs pre-trained BERT as a sentence encoder and directly generates sequence-level embeddings for log templates.}

\noindent \textbf{Sequence-level representation.}
Usually, log data contains a sequence of log entries that can be sorted according to the chronological order indicated by timestamps. The whole log data can be grouped into a set of log sequences with different approaches (e.g., fixed windows, sliding windows, and session windows~\citep{he2016experience, chen2021experience, le2022log}\accepted{\heng{use a reference like the Loglizer paper}} according to the needs of downstream tasks. We call the embedding for this abstraction level as \textbf{sequence-level} representation. \response{R2.1}{Most of the traditional ML models (e.g., SVM, decision tree) work on the representations of this level~\citep{he2016experience}.}\accepted{\heng{give an example}} Sequence-level representation can be acquired by aggregating log event-level representations or by using sequential models (e.g., RNNs, Transformer).
%\Foutse{it would be nice to add some illustrations to explain these different levels of abstraction} \heng{use an illustrative figure to illustrate the different levels with a few lines of logs, refer to the figure while describing these different levels}

\subsection{Applications of log representations in automated log analysis}
%\secdescription{Describe how log representations are used in general log analysis (in an abstract way).Describe the downstream tasks used in your study.}

When log data is represented as vectors or other structured data structures, various automated log analysis models can be built upon to realize various downstream tasks, such as anomaly detection~\citep{he2016experience, chen2021experience}, performance modeling~\citep{liao2020using}, predictive analysis~\citep{katkar2014use}, or casual analysis~\citep{jarry2021quantitative}.   %. The representative downstream tasks are: anomaly detection, predictive analysis, causal analysis, log search and etc. 
%\heng{Explain a concrete example}.

Anomaly detection is the most representative downstream task of log analysis~\citep{fu2009execution, xu2009detecting, he2016experience, chen2021experience, meng2019loganomaly, le2021log, wang2018anomaly, zhang2019robust, du2017dl, lu2018detecting, nedelkoski2020self}. Log-based anomaly detection approaches identify anomalies inside a log sequence according to the occurrence patterns of log events. %, parameters in the logs, log levels and etc. 
With log representation, the anomaly detection task can be formulated as both unsupervised and supervised methods. Unsupervised methods adopt unsupervised machine learning algorithms (e.g., isolation forest) to mine the normal patterns of log data usually with the hypothesis that anomalies are unusual in the data sequence~\citep{liu2012isolation}, while the supervised methods usually treat anomaly detection as a classification problem and employ classifiers (e.g., decision tree) to learn the normal and abnormal modes~\citep{he2016experience}.

Based on the availability of well-annotated datasets and the richness of related works, this work adopts the log-based anomaly detection task as our protocol to study the log representation. According to prior works~\citep{he2016experience,chen2021experience}, supervised anomaly detection models usually achieve better performance and have better stability across datasets than their unsupervised counterparts. Also, the performance of unsupervised models is sensitive to their hyper-parameters. According to our experiments, unsupervised models favour different hyper-parameters when working on different datasets and need manual tuning. Otherwise, they may generate inferior results that will influence the comparisons of log representations. Based on these observations, we only focus on supervised models in this work to eliminate interference from these factors.

% Our work only focus on anomaly detection task, as it's

%\xingfang{very few papers on the following topics.}

%The predictive analysis~\citep{katkar2014use} aims to predict the timing and ordering of future events with log data. It is similar to anomaly detection for many predictive schemes which also adopted classifiers to predict the probability score for the appearance of possible future events. A suitable and effective log representation can provide accurate and sufficient information for this application scenario \Foutse{can you explain more or provide some references?}. 

%The casual analysis~\citep{jarry2021quantitative} intends to trace the cause of failures in a software system. The causality can be measured with the ability to predict a future log with a previous log \Foutse{can you elaborate? with example possibly?}.

%The log search is a practical application that can help software practitioners find and locate history logs with their inputs \Foutse{can you explain a bit the use case?}. Nearest neighbour search is one of the algorithms that can be used to generate query results based on log representations.

\section{Related Work} \label{sec:related}

%\secdescription{Discuss related work and group them into a few categories.}

%A typical automated log analysis framework follows the following steps. Log parsing is usually the first step to preprocess the semi-structured log data to extract specific information related to a specific task. Log mining often consists of extracting and representing features of the log data and a follow-up processing for the corresponding tasks. 
%As the most representative task in log analysis, anomaly detection has been well studied and different feature extracting, and various log representation approaches have been applied to this task. So, our survey here will focus on representation techniques in existing anomaly detection frameworks. 
In this section, we discussed existing log representation techniques and their applications in log analysis tasks.
Generally, existing log representation techniques can be classified into two categories based on the mechanism to generate log representation: the classical approaches based on handcrafted features and semantic-based approaches. 
In addition, we discuss prior art on anomaly detection which is our focused downstream task in this work.%\heng{to add}

%\heng{use consistent wording with the studied log representation techniques.}

%\subsection{Traditional feature engineering for log representation}
\subsection{Classical log representation techniques and their applications}

There are several kinds of features manually designed by researches according to their domain knowledge to represent log data.

\paragraph{\textbf{Log template ID\accepted{\heng{(event level)}}}} As log data is sequential and log messages are generated by a limited amount of logging statements, a log sequence can be easily presented as a sequence of log template id (a.k.a. log key) after being parsed with a log parser~\citep{zhu2019tools}. Although this approach ignore a lot of information from logs, it is an effective representation that reflects occurrence patterns of log templates in a log sequence. %Therefore, anomalies may be spotted out with abnormal occurrence pattern of log templates in anomaly detection.
\response{R2.1}{Log template ID is an event-level representation, which can be aggregated into Message Count by a count vectorizer.}

%\heng{Try to merge similar work / work in the same category together and just use one example to explain. This paragraph is an example.}
Prior works use log template IDs to detect anomalies, as anomalies may be spotted out with abnormal occurrence pattern of log templates~\citep{du2017dl, lu2018detecting}. 
For example,~\citet{du2017dl} proposed the DeepLog anomaly detection framework, in which a sequential anomaly detection model is trained with log keys. Combined with another performance anomaly detection model, the framework achieved the state-of-the-art detection performance at the time it was proposed.

%Like DeepLog, Lu et al.\citep{lu2018detecting} proposed LogCNN, which consists of logkey2vec embeddings, a trainable layer that can be optimized with gradient descent to present log templates by maping each log key into a vector, and thus a log sequence is presented as a matrix. Then, the feature matrix is input into the CNN network, which consists of 1D Convolutional layers, dropout layers, max-pooling, and a softmax layer to produce the probability distribution results. 

\paragraph{\textbf{Message Count}} Unlike log template ID representation, Message Count (a.k.a. event count, log count, log message counter) Vector counts the occurrences of log templates in a log sequence and the length of representation depends on the amount of log templates in a whole log data, and thus is unrelated to the length of the log sequence. \response{R2.1}{ Message Count is a sequence-level representation.}

%\heng{merge the paragraphs below into one short paragraph: Prior works also use message count in a session to detection anomalies [citep all the mentioned papers]. For example, [then give one single eample]}

It is one of the most common traditional log representation approach that adopted by various log analysis frameworks. For example,~\citet{he2016experience} use the log count as features and fed them a logistic regression model to detect anomalies.~\citet{xu2009detecting} adopt the unsupervised dimension reduction method PCA with event count matrix to detect anomalies.~\citet{lou2010mining} input event count to invariant mining algorithm to detect anomalies.

\paragraph{\textbf{TF-IDF}} is a commonly used weighting technique in information retrieval and data mining. For log data, TF-IDF weighting can be either used to weight values in Message Count Vector or serve as a feature itself to present tokens in a log entry. For example,~\citet{wang2018anomaly} use the TF-IDF values of tokens in a log event to form the feature vectors. 
Some researchers modified TF-IDF to better suit the characteristics of log data. For example,~\citet{meng2021logclass} apply the popular bag-of-words model to generate embedding and design the Inverse Location Frequency (ILF) method (a modified version of IDF\citep{salton1988term} designed for logs) to weight the words of logs in feature construction. \response{R2.1}{When TF-IDF operates on tokens within log events, it produces representations at the event level. Alternatively, when it processes the sequence of template IDs, it generates representations at the sequence level.}

%Different from previous works, Lin et al.\citep{lin2016log} further revised the event count vector representation with two methods to weight log events: IDF-based and contrast-based event weighting. For each log event, IDF-based event weighting calculates the IDF( Inverse Document Frequency)\citep{salton1988term} values.

\paragraph{\textbf{Combined features}} Also, there are other works that try to combine different features and representations for log data.~\citet{liang2007failure} proposed a failure prediction model for log data generated from IBM Blue Gene/L. In this work, six groups of features are generated, including the number of events of different severity, event distribution, inter-failure times, and so on. \response{R2.1}{These sequence-level representations} are further processed by four classifiers(e.g., SVM, KNN) for later anomaly prediction.

\subsection{Semantic-based log representation techniques and their applications}
%\accepted{\heng{Better to category them into studied semantic log representation techniques: Word2Vec-based, FastText-based, BERT-based, others. For exmaple, Word2Vec-based does not need to be exactly word2vec, but any that is related to or based on word2vec.}}

Unlike classical approaches, semantic-based approaches employ deep-learning techniques that do not rely on manually designed features. As logs are semi-structured texts and log messages contain semantic information, some studies leveraged deep learning techniques in natural language processing and information retrieval to represent and analyze log data.

\paragraph{\textbf{Static Embedding}} Some works are inspired by static word embeddings, which have been demonstrated to be more effective than log keys and log count. \response{R2.1}{Static embedding techniques create embeddings for tokens in log events, resulting in token-level embeddings that can be further aggregated into higher-level embeddings.}~\citet{meng2019loganomaly} proposed a log representation approach named Template2Vec. By embedding the log template with dLCE~\citep{nguyen2016integrating} to a vector, this approach presents the first step towards considering semantic and syntax information in log data.

The subsequent study proposed Logsy \citep{nedelkoski2020self}. In this work, two operations are applied to input tokens: token embedding and positional encoding. Before being embedded into vectors, log messages are split into word tokens and numerical characters and commonly used English words are removed. Then, these vectorized tokens are input into the subsequent encoder of the Transformer \citep{vaswani2017attention} module with multi-head self-attention. 

%Meng et al. further proposed Log2Vec \citep{meng2020semantic} to address two challenges template2Vec faced by integrating the Log-specific word embedding method (LSWE) with an OOV word processor. The first is that template2Vec may not capture the precise meanings of domain-specific words in logs. The second is the lack of a mechanism to handle out-of-vocabulary(OOV) words. They applied Log2Vec to log classification and anomaly detection tasks which demonstrate the approach can extract critical features from the log and improve the performances of downstream tasks. 

\citet{zhang2019robust} proposed a log-based anomaly detection approach called LogRobust. They leverage Drain \citep{he2017drain} to obtain log templates and encode log templates with a pre-trained FastText model combined with TF-IDF weight. Then, an attention-based Bi-LSTM model is used for anomaly detection. With semantic embeddings, It can identify unstable log events with similar semantic meanings. 

\paragraph{\textbf{Contextual Embedding}} \citet{le2021log} proposed NeuralLog, which does not rely on any log parsing. In NeuralLog, each log message is directly transformed into semantic vectors after removing numbers and special characters. A pre-trained BERT model is employed to encode log messages into a fixed-dimension vector representation. \response{R2.1}{Similar to static embeddings, contextual embeddings can operate at the token level. Nonetheless, pre-trained models may also generate event-level embeddings using their unique structures, such as the pooler layer in BERT~\citep{devlin2018bert}.}

\accepted{\heng{Does this belong to semantic-based representations?}\xingfang{no. it was a mistake.}}
\subsection{Graph-based log representation techniques and their applications}
\response{R2.9}{Recently, a group of studies introduced Graph Neural Networks (GNNs)~\citep{xie2022loggd, wan2021glad} to log anomaly detection. Unlike previously-mentioned \accepted{\heng{previously-mentioned?}} approaches, which mainly utilize the sequential or quantitative patterns of log events in log sequences, GNN-based methods transform log sequences into graphs and leverage the spatial structural relationships among logs. \response{R2.1}{Typically, these methods generate representations at the sequence level by utilizing features from lower levels.} Some previously mentioned representations can be incorporated into the graph structure as features of nodes and encoded by a GNN-based graph encoder. \accepted{\heng{this sentence is not clear. Maybe: Previously-mentioned representations can be incorporated into a graph-based representation through a GNN-based graph encoder?}} The experiments show that these approaches achieved promising results and robustness against the variation of window size. As the representation learning process is linked to downstream tasks, we are unable to incorporate this type of representation into our experimental framework. Therefore, it is not included in our experiments.}
\accepted{\heng{make sure we justified why we did not consider this type of representation}}

\subsection{Anomaly detection}

% \heng{Discuss prior work on anomaly detection and explain our difference from them: we perform a comprehensive evaluation of the impact of log representations on anomaly detection.} \xingfang{fixed}

% As a highly in-demand task of automated log analysis, anomaly detection approaches have been widely studied in the last decades~\citep{fu2009execution, xu2009detecting, he2016experience, chen2021experience, meng2019loganomaly, le2021log, wang2018anomaly, zhang2019robust, du2017dl, lu2018detecting, nedelkoski2020self}. 
% \response{R3.2}{Due to the fact that many automated log analysis tasks share similar, if not the same, frameworks that process log data, we hope our work can also inspire and support the designs of workflows of other tasks, despite the fact that only one downstream task is being evaluated in our work. We further discussed these in the background and our threat to validity.}

As one of the most studied downstream tasks in the domain of automated log analysis, anomaly detection aims to detect abnormal system behaviours to help developers and operators uncover system issues and solve anomalies. Log data is a good source of information that can be utilized for anomaly detection models to evaluate the status of a system, as it may contain the indexes of the availability of system resources and the running status of services. The log sequence can also reflect the execution paths of a system. From these pieces of information, potential failures or unusual execution sequences can be spotted according to the regular pattern. Therefore, as a highly in-demand task of automated log analysis, log-based anomaly detection has been widely studied, and various approaches have been developed in the last decades.

Traditionally, developers may check system logs with keywords or use rules to find anomalies and locate the bugs in systems with their domain knowledge. Manual inspections are erroneous and unstable for large software systems that generate tons of logs in a short period. Rule-based approaches demand the manual construction of rules and can not adapt to fast-evolving software systems. Therefore, machine learning is adopted in many log-based anomaly detection approaches.
% Among these approaches, supervised learning methods usually achieve superior performances~\citep{he2016experience, chen2021experience}. 
In this study, we only focus on supervised learning methods, as we discussed in Section~\ref{sec:background}. Supervised anomaly detection is defined as a machine-learning task of deriving a classifier with an annotated log sequence. The annotations mark the normal or anomalous states of log sequences or log events. Here, we list the most representative related works that utilize supervised learning methods in anomaly detection.

\paragraph{\textbf{Traditional Methods}} Most of the traditional machine learning methods adopt message count vector as their log representation approach. A training instance for traditional models usually consists of an event count vector for a log sequence and its corresponding label. With training instances, classifiers can be trained to classify new instances. Logistic regression is a statistical model that is widely used in anomaly detection. It estimates the probability of normal and anomalous according to the input vector. The decision tree is a tree-based model that is constructed in a top-down manner with training data. Each node presents a split of an attribute with the criteria of information gain. The decision tree was also applied to log analysis in previous works~\citep{chen2004failure}. Event count vectors are used to construct the decision tree, and predictions for new instances are given with tree structure. Support Vector Machine (SVM) is a common supervised method for classification. A hyperplane is constructed by maximizing the distance between the hyperplane and the closest point(s) of different classes to separate instances in high-dimension space. SVM was employed to detect failures~\citep{liang2007failure} with statistical features of occurrences of log events.

\paragraph{\textbf{Deep Learning Methods}} Different from traditional methods, the input feature of deep learning methods for log anomaly detection varies greatly. The most basic model is based on Multi-layer Perception (MLP). MLP is a feed-forward structure that maps a set of input vectors to a set of output vectors. MLP model serves as a baseline model for log-based anomaly detection in previous works~\citep{lu2018detecting}. Convolutional Neural Networks (CNNs) were first adopted for log anomaly detection by~\citet{lu2018detecting}. This work uses convolutional layers containing different kernels to extract features from vectors generated with a codebook that maps the logs to embedding vectors. Long Short-Term Memory (LSTM) is commonly used for mining the patterns from log data in many automated log analysis frameworks~\citep{du2017dl, meng2019loganomaly}. However, the mechanisms of prior works vary: some works~\citep{du2017dl} used log template ID as input, and LSTM is used to learn the occurrence patterns of log templates in normal and abnormal log sequences, while there is another line of works that take embedding vectors of log templates as input~\citep{meng2019loganomaly}. \response{R2.2.}{
Transformer-based models have been applied in the log-based anomaly detection task by some recent works~\citep{nedelkoski2020self, le2021log}. The transformer blocks in these models\accepted{\heng{transformer-based models?}} can capture contextual information from input sequences with the self-attention mechanism. These models exhibit promising results in log-based anomaly detection tasks. However, the previous works utilized transformer-based models with different formulations of the log-based anomaly detection task. For example, Logsy~\citep{nedelkoski2020self} formulates the anomaly detection problem as discrimination between normal logs from the system of interest and auxiliary logs from other systems, in which anomalies are detected based on only their log messages and sequential information is ignored. \accepted{\heng{would be good to be more specific, but be brief}}The best practices for using transformers in log analysis are \accepted{\heng{are}}still unclear. Therefore, we do not include this sort of method in our experiments.}\accepted{\heng{did we explore it?}\xingfang{no}}

\section{Experimental Design} \label{sec:design}

%\secdescription{Describe how you design your experiments, including selection and implementation of log representations, downstream tasks and datasets, models, how you implement the downstream tasks (e.g., parameters used, how to apply the log representations), evaluation methods.}

In this section, we introduce the design of our experiments to evaluate the effectiveness of different log representation techniques by assessing their impact on the performance of selected automated log analysis tasks. First, we give an overview of the experiment workflow. Then, we discuss log representation techniques studied in the experiments %are discussed, which are used 
to generate feature vectors for follow-up downstream tasks. Then, our selected downstream tasks with their corresponding models are introduced. We also review the metrics that we use to evaluate the performances of each studied downstream task.

As our focus is on the impact that log representation techniques have on the performances of downstream tasks, we select the most representative automated log analysis task (i.e., anomaly detection) and combine different models of it with different representation techniques. %and different configurations\heng{is different configurations studied?} of the representation methods. 
Aside from the comparative evaluation of studied representation techniques on follow-up models, we also look into the log parsing and feature aggregation process that can affect the effectiveness of log representations. %\accepted{\Foutse{can you be more specific?}}% centering on the research questions we propose. 

\subsection{Overview}
%\secdescription{Provide an overview figure and corresponding text.}

\begin{figure*}[!t]
    \centering
	\includegraphics[scale=0.53]{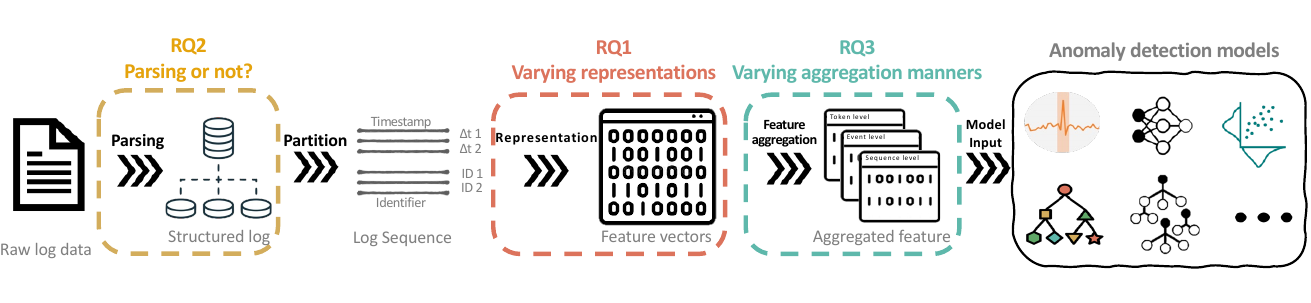}
	
	\vspace{-3mm}
	\caption[Workflow]{General workflow of our experiments. The variations for each research question are highlighted with dotted boxes. %\accepted{\heng{does it make sense to change ``feature extraction \& representation'' (a typo there) to ``Log representation?''} \heng{use the figure to highlight the variations that you perform in each RQ: RQ1 varies the representation techniques; RQ2 varies parsing; RQ3 varies aggregation. Highlight the three variations in this figure and label them with RQ1 (varying representation techniques), RQ2 (parsing or not), RQ3 (varying aggregation methods) respectively. So people can understand your RQs by looking at the figure. }}
% 	\accepted{\heng{change ``Input'' to ``Model input''}}
    \accepted{\heng{Text not clear. Make the text darker or bigger }}
	}
	\vspace{-2mm}
	\label{fig:workflow}
\end{figure*}

Figure~\ref{fig:workflow} shows the general workflow of our experiments. Raw log data is unstructured textual data. As most log representation techniques require structured log data as inputs, a log parsing step is often applied to obtain structured log data. In this work, the Drain parser~\citep{he2017drain} is adopted in our experimental workflow, as it is shown to have superior parsing performances on most of the datasets~\citep{zhu2019tools}. 
However, log parsing may not be needed for some log representation techniques. For example, log parsing process is deserted in NeuralLog~\citep{le2021log}, a recent anomaly detection workflow that achieved results outperforming the other existing approaches.
%\accepted{\heng{briefly give an example with reference}}
%For other representation techniques that demand raw log message as input, parsing process is ignored and sometimes some other preprocessing processes are adopted to remove some stop words and non-character tokens, such as delimiters, operators, punctuation marks, and number digits. 
Then, log data is fed into representation algorithms to get numerical representations of different abstraction levels. According to the level of abstraction of the log representation, different models of downstream tasks are selected to mine critical information from the data according to the specific task and yield the analytical results. 
% Finally, metrics are adopted to evaluate the performances of the models that use different log representations as the inputs.

\subsection{Studied log representations}
%\secdescription{Describe the studied log representations, why select them. Describe how you implement them (in plain language).}

As our goal is to conduct an evaluation of different log representation techniques, we selected the most representative techniques for the study. We implemented the studied techniques following the common practices of previous works  ~\citep{chen2021experience,he2016experience} to better compare their characteristics and quality. However, different previous works have different implementations with minor alternations for some representation techniques. We adopt and synthesize these open-source codes or implement them ourselves from scratch. Existing log representation techniques can be classified into two categories: classical and semantic-based approaches. 

\noindent\textbf{Classical log representation techniques.} %Traditional approaches usually represent logs with handcrafted features or some statistical characteristics of the occurrence of tokens or log templates in a log sequence. 
%\heng{briefly explain an example with reference}
%However, the deep-learning-based approaches usually treat log messages as natural languages and involve a language model to convert tokens in the log message into embedding vectors. This section gives a brief introduction to our studied log representation techniques.
%\xingfang{moved forward.}\heng{use more structured format (bullets points or a table (better)) to describe each selected representation technique and provide reference. Even better, use a running example to illustrate these representations. It is important to let readers understand these representations clearly as they are key of the paper.}
For classical representation techniques, we select message count vector, template ID-based TF-IDF (TF-IDF ID) and text-based TF-IDF (TF-IDF Text) feature representation. The message count takes log template indexes as input. It presents a log sequence with a vector counting the event occurrences from each log template. The event template ID-based TF-IDF (TF-IDF ID) weighting weights each event template ID with their respective TF-IDF value. In template text-based TF-IDF (TF-IDF Text) representation, we used the TF-IDF values of tokens in the template of a log event to represent a log message. For a sequence, we calculate the average of feature vectors of its log events to form the representation for the sequence.

\noindent\textbf{Semantic-based log representation techniques.}
%Deep-learning-based approaches usually treat log messages as natural languages and involve a language model to convert tokens in the log message into embedding vectors. 
%\heng{briefly explain an example with reference}
%This section gives a brief introduction to our studied log representation techniques.
%\heng{use more structured format (bullets points or a table (better), can be merged in the same table with the classical ones) to describe each selected representation technique and provide reference.}
For semantic-based log representation techniques, we choose three commonly adopted techniques in existing automated log analysis frameworks as our objects of study: Word2Vec, FastText, and BERT. For each of them, we leverage the pre-trained models trained with natural language corpus as related works do~\citep{zhang2019robust,le2021log}. For Word2Vec, we use the word vectors generated by the model pre-trained with Google News dataset\footnote{\url{https://code.google.com/archive/p/word2vec/}}. Pre-trained Word2Vec can generate many out-of-vocabulary~(OOV) words when processing log data and is unable to handle them in a proper way. So, we assign the zero vector for OOVs when generating Word2Vec representations for the studied datasets. For FastText, we leverage the off-the-shelf word vectors, which were pre-trained on Common Crawl Corpus and Wikipedia~\citep{grave2018learning}. FastText can handle OOV words by summing up embeddings for its component character-n-grams. Therefore, FastText is able to generate embeddings for OOV words in logs, although the embeddings may not be effective. For BERT~\citep{devlin2018bert}, we utilize the pre-trained base model~\citep{turc2019}. The sentence embedding is generated by the second-to-last encoder layer of the model, which is 768 dimensions. The second-to-last hidden layer is chosen as the last layer is too close to the target functions during pre-training, which may contain biases. 

\subsection{Downstream models and datasets}

\subsubsection{Anomaly detection models and implementations}
%\heng{A little reoganization needed: First summarize what models you choose (traditional models and deep-learning models) and explain why choosing them, then go to the individual models and their configurations, finally how you implement them (e.g., based on Loglizer, use PyTorch etc.}.
%\accepted{\heng{describe such implementation details at last}}

%And for each model, we may adjust some hyper-parameters according to the dimension of input features to avoid extreme poor performances, as the dimensions of feature vectors of the same representation techniques may vary according to the datasets.

%\accepted{\heng{Also need to describe traditional models. They can be short if space does not allow.}\xingfang{as these models are well-known, i think we dont need to elaborate here.}}
%\accepted{\heng{describe such implementation details at last}}
% For the three studied deep-learning models, we implement them from scratch using the PyTorch framework. 
%As our goal is to evaluate the effectiveness of log representation techniques rather than pursue high performances of downstream models,
%\accepted{\Foutse{this argument is risky...because if models are underspecified then we may not be making fair evaluations!}\xingfang{I rephrased it to be less risky.}}
%we keep our implementations simple while trying to maintain consistencies (e.g., network structures, hyper-parameters) to the original papers of these models.
We select 7 supervised machine learning anomaly detection models to evaluate the studied log representation techniques. SVM, decision tree, logistic regression, and random forest are traditional machine learning models. These models are commonly used and well-studied in various application scenarios and often serve as a baseline in automated log analysis tasks~\citep{he2016experience}. For deep-learning models, we choose MLP, CNN, and LSTM models. The MLP model is selected as a baseline for log-based anomaly detection in prior work~\citep{lu2018detecting}. CNN and LSTM are widely employed in many automated log analysis frameworks~\citep{lu2018detecting,du2017dl,meng2019loganomaly}.

We employ these well-studied machine learning models based on the fact that they are commonly adopted in anomaly detection workflows or other automated log analysis frameworks. By selecting these widely adopted models, we believe that the findings of our work may stand a better chance of generalizing to other log-related automated analysis tasks. We briefly introduce the implementation of the studied models in the following, while details can be found in our replication package\footnoteref{reppack}.

% \Foutse{can you add the reference to the replication package here?}

\paragraph{Traditional models} For traditional anomaly detection models, we follow the implementations of Loglizer~\citep{he2016experience}. However, their implementations only take the event count matrix generated with session windows as input. In our case, the input dimensions vary according to the studied log representations. Same as Loglizer, all of our studied traditional models take sequence-level representation as input. We modify hyper-parameters of these models according to the input dimensions of our generated log representations. 

\paragraph{Multi-layer Perception (MLP)} 
%MLP is a feed-forward structure that maps a set of input vectors to a set of output vectors. Here, 
We follow the similar implementation of the baseline model in~\citep{lu2018detecting}, we treat the anomaly detection task as a binary classification problem and use a MLP with one hidden layer with 200 neurons as a binary classifier. The inputs are feature vectors of different log representation techniques, and the outputs are the one-hot encoding of the binary labels. Cross entropy loss is used as the criterion to train the three-layer network. MLP also takes sequence-level log representation as input.

\paragraph{Convolutional Neural Network (CNN)} In our work, we implement exactly the same network structure as in the original work~\citep{lu2018detecting}. However, instead of taking log keys as input and using the codebook to map log keys into embeddings, our network substitutes the codebook with a fully-connected layer with 50 neurons, which maintains the same embedding size as the original work for the convolutional layers to process. Network details can be find in our replication package.
% are listed in the table~\ref{tab:cnn_params}. 
The CNN models require event-level log representation as input, and demand the input sequence are of same length. As the sessions of log data may contain different numbers of log messages, we sliced the sessions with a sliding window.

\paragraph{Long Short-Term Memory (LSTM)} 
%LSTM is a special type of the recurrent neural network (RNN) structure that can track long-term dependencies in input sequences. As it has the ability to learn and predict sequential data and log data usually follows a chronological order and execution sequence, 
% LSTMs are commonly used for mining the patterns from log data in many automated log analysis frameworks~\citep{du2017dl,meng2019loganomaly}. However, the mechanisms of prior works vary:
%Different works take different feature representation as inputs and try to mine different information from the log sequences with the models. %\accepted{\Foutse{consider breaking the sentence in two parts, it is too long!}} 
% Some works (e.g.,~\citep{du2017dl}) used log template ID as input, and LSTM is used to learn the occurrence patterns of log templates in normal and abnormal log sequences, while there is another line of works that take embedding vectors of log templates as input~\citep{meng2019loganomaly}. 

There are different mechanisms of using LSTM to detect anomalies in log sequence in prior works. As the aim is to compare different log representations, our implementation treats different log representations as the input feature of the LSTM model. Similar to CNN models, LSTM models require fixed length event-level representation as input. Network details can be found in our replication package\footnoteref{reppack}.
% \Foutse{add the reference!}
% are listed in Table~\ref{tab:LSTM_params}.

% \begin{table}[htb]
% \centering
% \caption{Network details of our LSTM implementation}
% \label{tab:LSTM_params}
% \begin{tabular}{|c|c|c|ll} 
% \cline{1-3}
% \begin{tabular}[c]{@{}c@{}}Layer\end{tabular} & Parameters                      & Output                &  &   \\ 
% \cline{1-3}
% \textbf{Input }                                 & ~$[win\_size * Embeddin\_size]$ & N/A                   &  &   \\ 
% \cline{1-3}
% \textbf{LSTM}                                   & Hidden\_dim = 8                 & $Embedding\_size * 8$ &  &   \\ 
% \cline{1-3}
% \textbf{FC}                                     & $[8 * 2]$~                      & $2$                   &  &   \\ 
% \cline{1-3}
% \textbf{Output }                                & Softmax~                        &                       &  &   \\
% \cline{1-3}
% \end{tabular}
% \end{table}

\subsubsection{Datasets and preparations}
\label{sec:data_prep}
Our experiments evaluate the existing representations with the following four public log datasets provided by LogHub~\citep{he2020loghub}:
%(a \heng{X}G HDFS log dataset~\citep{xu2009detecting} and a \heng{Y}M BGL log dataset~\citep{oliner2007supercomputers}) that are widely studied in existing works~\heng{citep papers using these two datasets}. %, for better comparison with other works.
%\xingfang{X,Y mean the sizes of the dataset? i think it would be better to use the amount of log messages rather than size? as the size will change according to the format(compressed/json/structured log).}

\noindent $\bullet$ The HDFS dataset ~\citep{xu2009detecting} is collected from the Amazon EC2 platform. It contains more than 11 million log events, and each event is associated with a block ID, by which we slice log data into a set of sessions, which are the sub-sequences of the entire log sequence. For each session, labels are given to indicate whether there exist anomalies. There are a total of 575,061 log sessions with 16,838 (2.9\%) anomalies.

\noindent $\bullet$ The BGL dataset ~\citep{oliner2007supercomputers} is recorded from the Blue Gene/L (BG/L) supercomputer system at Lawrence Livermore National Labs (LLNL) with a time span of 215 days. This dataset contains 4,747,963 annotated log messages, where 348,460 (7.3\%) are labelled as failures. Unlike HDFS, log messages in BGL do not have identifiers for separating logs from different job executions, processes or threads. So, grouping techniques (e.g., time-based, fixed window-based, etc.) are adopted to form sub-sequences. For uniformity, we also call these sub-sequences in BGL as sessions. 

\response{R1.3, R2.4, R3.1 begin}{}\accepted{\heng{color the changes}}

\noindent \pararesponse{$\bullet$ The Spirit dataset is also a well-used public log dataset~\citep{oliner2007supercomputers}, which Sandia National Labs collected from their Spirit supercomputing system. There are more than 272 million log messages in total.
%, among which around 172 million are labelled as anomalies\heng{maybe do not mention the 172m anomalies as it is not important. Only the number of anomalies in the sample is important}. 
As the whole dataset is too large for us to process, we use a subset containing the first 5 million log messages in our work, which follows the practice of prior work~\citep{le2022log}. In the subset, 15.5\% of the log messages are marked as anomalies. The subset is shared in the replication package.
\accepted{\heng{mention that the subset is shared in the replication package for replicability}}}

\noindent \pararesponse{$\bullet$ The Thunderbird dataset~\citep{oliner2007supercomputers} is also a public log dataset from Sandia National Labs. There are around 211 million log messages in total
%, with more than 3 million\heng{3m not important here} logs manually identified as abnormal. 
We followed the practices of previous works~\citep{le2021log, le2022log} and extracted a continuous chunk of 10 million\accepted{\heng{a continuous chunk of 10 million}} log messages from the whole dataset, among which 4.1\% are labelled as anomalies. We also share the subset in our replication package.}

\accepted{\heng{People may question why we did not choose the beginning chunk as for the Spirit dataset. Maybe we can say this:? We purposefully did not consider the first 10 million log messages, because the beginning part of this log dataset is filled with a very large portion of anomalies caused by the instability of the system at the beginning.}
\xingfang{The problem of the first 10m is unclear. Maybe there are a lot of mistakes in their annotation. If we mention the percentage of the anomalies, it seems the models can not handle the unbalanced cases.}}
\accepted{\heng{mention that the subset is shared in the replication package for replicability}}

\response{R1.3, R2.4, R3.1 end}{}

According to the common practices ~\citep{chen2021experience,he2016experience} of dataset preprocessing and grouping, we prepared the studied datasets with the following configurations:

% \heng{bring Table~\ref{tab:Sliding_params} here, also explain the special treatment of the CNN and LSTM (time windows) after the general sessions.}

\begin{table}
\centering
\caption{Grouping techniques and default window size settings for studied datasets \accepted{\heng{This table is not only for sliding window configuration? If the table is updated then color the title to mark the change }\xingfang{forgot to change the title}}
% \heng{make it clear what is common to all models and what is unique for specific models (like CNN, LSTM); split it into two if needed.
}
\label{tab:Sliding_params}
\resizebox{\linewidth}{!}{
% \begin{tabular}{lcccccccc} 
% \hline
% \multirow{2}{*}{Dataset} & \multirow{2}{*}{\begin{tabular}[c]{@{}c@{}}Splitting\\Criterion\end{tabular}} & \multicolumn{2}{c}{\# of sessions} & \multirow{2}{*}{\begin{tabular}[c]{@{}c@{}}Ave. \# of log\\~per session\end{tabular}} & \multirow{2}{*}{\begin{tabular}[c]{@{}c@{}}Window\\size\end{tabular}} & \multirow{2}{*}{Stride} & \multicolumn{2}{c}{\# of windows}  \\ 
% \cline{3-4}\cline{8-9}
%                          &                                                                               & Train   & Test                    &                                                                                       &                                                                       &                         & Train   & Test                     \\ 
% \hline
% \multicolumn{1}{c}{HDFS} & Session ID                                                                    & 402,542 & 172,519                 & 21.87                                                                                 & 30                                                                    & 1                       & 549,991 & 177,276                  \\
% \multicolumn{1}{c}{BGL}  & Time (6h)                                                                     & 575     & 143                     & 6,564.61                                                                              & 50                                                                    & 50                      & 72,367  & 22,230                   \\
% \hline
% \end{tabular}
\begin{tabular}{ccccccc} 
\hline
\multirow{2}{*}{Dataset} & \multirow{2}{*}{\begin{tabular}[c]{@{}c@{}}Grouping\\Criterion\end{tabular}} & \multicolumn{2}{c}{\# of sessions} & \multirow{2}{*}{\begin{tabular}[c]{@{}c@{}}Ave. \# of log\\~per session\end{tabular}} & \multirow{2}{*}{\begin{tabular}[c]{@{}c@{}}Window\\size\end{tabular}} & \multirow{2}{*}{Stride}  \\ 
\cline{3-4}
                         &                                                                              & Train   & Test                     &                                                                                       &                                                                       &                          \\ 
\hline
HDFS                     & Session ID                                                                   & 402,542 & 172,519                  & 22                                                                                    & 30                                                                    & 1                        \\
BGL                      & Time (6h)                                                                    & 575     & 143                      & 6,565                                                                                 & 50                                                                    & 50                       \\
Spirit                   & Time (1h)                                                                    & 938     & 235                      & 4,208                                                                                 & 50                                                                    & 50                       \\
Thunderbird              & Line (100l)                                                                  & 79,773  & 19,944                   & 100                                                                                   & 30                                                                    & 10                       \\
\hline
\end{tabular}
}
\end{table}

\noindent \textbf{Preparation for the HDFS dataset.}
For the HDFS dataset, as the available annotation labels are based on blocks ID, an identifier that marks the different execution sequences, we use it as the clue to group logs into sessions.
%and use the block ID as the unique identifier to annotate\heng{distinguish? be careful label usually means anomaly or normal labels} sessions \heng{reference}. 
We use 70\% of the sessions as the training set and the other 30\% as the test set by following the common practices of dataset splitting in supervised learning tasks (e.g., \cite{el2017learning,lyu2021empirical}). During splitting, we shuffle the sessions while maintaining the time-based sequence of log messages inside each session~\citep{chen2021experience}.
Recent work~\citep{le2022log,lyu2021empirical} suggests that the random shuffling process can cause data leaking problems. However, as the main focus of our work is the impact of log representation rather than the performance of the downstream models, the random shuffling process will not undermine our evaluation.

\noindent \textbf{Preparation for the BGL dataset.}
For the BGL dataset, we do not have identifiers to separate the log items into different execution sequences. So, we choose to \response{R3.3}{group} the log messages according to the timestamp. We refer to the \response{R3.3}{grouping} approaches of prior papers that adopt the BGL dataset and group the log messages with a fixed window of 6 hours~\citep{he2016experience}. After the time-based \response{R3.3}{grouping}, there are 718 sessions. As the number of sessions is far less than that of the HDFS dataset, we use 80\% of the sessions as the training set and 20\% as the test set instead of a 70\%/30\% splitting, following the practices in prior work~\citep{chen2021experience,le2021log,meng2019loganomaly}. Similarly, we shuffle the sessions while maintaining the time-based sequence within each session. The labels are merged from that of the log messages inside each session. If any of the log messages inside a session is labelled as an anomaly, the whole session is recognized as an anomaly, following the approach used in prior work~\citep{he2016experience}.

\response{R1.3, R2.4, R3.1 Begin}{}\accepted{\heng{color the changes}}

\pararesponse{\noindent \textbf{Preparation for the Spirit dataset.} Similar to the grouping configuration of the BGL dataset, we group the log messages according to their timestamps. However, we adopt a fixed window of one hour instead of six hours following the configuration in prior work~\citep{le2022log}. After grouping, we get 1,173 sessions, with 221 anomaly samples. We further shuffle and partition the sessions into the training and test sets with an 80\%/20\% splitting.}

\pararesponse{\noindent \textbf{Preparation for the Thunderbird dataset.} Instead of adopting a one-hour fixed-window grouping, we employed a fix-length grouping to the Thunderbird dataset, as we noticed that the logs were unevenly distributed in time. If the sessions are grouped by a fixed-length time window,
% in some sessions grouped by fix-time window
\accepted{\heng{could not understand: If the sessions are grouped by a fixed-length time window?}}the number of logs in some sessions may be extremely large. We chose a window size of 100 lines, which is also a setting employed in experiments from previous work~\citep{le2022log}. After grouping, we get 99,717 sessions in total, among which 33,526 sessions are anomalies. We performed a sequential split of the sessions using an 80\%/20\% ratio to obtain the training and test sets. This approach helped improve the generalizability of our findings and establish their validity across all data selection configurations.}

\accepted{\heng{you did random shuffling for the other three dataset but sequential splitting for this data: any justification?}\xingfang{for increasing the generalization of our findings. To prove that they are tenable to all data selection configurations. We mentioned this somewhere later.}\heng{it should be mentioned here too, otherwise it causes confusion/question here. reviewers usually only focus on your highlighted changes.}}

\response{R1.3, R2.4, R3.1 End}{}

\accepted{\heng{color the changes like here}}
\response{R3.3}{
\noindent \textbf{Window size for sequential models}
As CNN and LSTM models require inputs to be of consistent sequence lengths, we need to further slice each log session with fix-length sliding windows. According to the characteristics of each dataset and the common practices in other works~\citep{chen2021experience}\accepted{\heng{references}}, we select the configurations of the sliding window in Table \ref{tab:Sliding_params} for the studied datasets as default settings. We further analyze the impacts of the variation of window size in RQ3.}\accepted{\heng{Mention that we analyze the impact of the window sizes in Section ...}}

\noindent \textbf{Log parsing}
%As our focus is to compare the characteristics of log representations, we want to minimize the influences of potential errors that log parsers will impose. 
Ideally, we would use a log parser that can convert the unstructured raw log data into structured log data without any error. In practice, however, existing log parsers cannot successfully parse all the log messages as the formats of log messages are usually diverse and complex. ~\response{R2.8 R3.6}{Continuous updates to existing parsing strategies and configurations are required due to new log templates and variations in log formats resulting from the evolution of software~\citep{zhang2019robust}.}\accepted{\heng{you can cite the LogRobust paper}}

\accepted{\heng{Never begin a sentence with also/and/but, see more here: https://www.merriam-webster.com/words-at-play/words-to-not-begin-sentences-with. Instead, use words like besides, in addition...}}

In fact, the impact of using different log parsers for automated log analysis has been explored in prior work~\citep{he2016evaluation}.~\response{R2.8 R3.6}{A recent work~\citep{le2022log} further investigated the impacts of data noise introduced by log parsing errors. The authors combined five anomaly detection models with four commonly-used log parsers and found that parsing errors induced by different parsers have distinctive impacts on downstream models. However, the patterns of the impacts remain to be explored.}

As our goal is to achieve the best possible parsing results to serve as input for our following processes, we do not compare the impacts of using different log parsers which may lead to different parsing results. According to~\citet{zhu2019tools}’s benchmarking work for log parsers, Drain~\citep{he2017drain} is the most accurate parser among their studied log parsers, which attains the highest accuracy on 9 out of 16 datasets. Therefore, we choose Drain as our log parser to preprocess the raw log into structured data and extract parameters from log messages in our work. % \xingfang{to pursuit the best parsing results}. 
However, the Drain parser can still generate a large number of 
%\heng{more than one thousand, or thousands of}
inaccurate templates for our studied datasets when we follow the default configuration indicated in the paper of~\citet{he2017drain}.
%\heng{make sure to fix citation format issues like this one (should use citet here and remove ``He et al.'' (duplicated)}
\response{R2.8 R3.6}{By examining the templates generated, it becomes apparent that certain parsing errors have occurred. For example, numerous log templates have been created with slight variations in certain fields that should be dynamic variables, but have instead been incorrectly identified as static text.\accepted{\heng{(i.e., they should be dynamic variables but were wrongly parsed as static template texts)}}\accepted{\heng{remove the first period}\xingfang{?}} To eliminate the impact of these inaccurate templates on our evaluation, we iteratively checked and appended the regular expressions designed for handling\accepted{\heng{handling these undetected dynamic variables}} these undetected dynamic variables\accepted{\heng{use ``variables'', ``areas'' not defined}}. We were able to decrease the number of resulting wrong templates. For example, after passing a set of regular expressions to the parser when parsing the Thunderbird dataset, the amount of log templates decreases from 2,241 to 1,488, in which many duplicate templates are removed. In a prior study~\citep{he2016evaluation}, a similar approach was utilized and it was verified that incorporating domain expertise (such as eliminating IP addresses) can enhance the precision of log parsing.} The details of the regular expressions can be found in our replication package.

%Other detailed operations on the datasets that are used to adapt to the input of each follow-up model will be introduced later in the approach sections of RQs \heng{revisit if this is still the case}.

\subsection{Evaluation methods}
%\secdescription{- Metrics used to evaluate the performance of the downstream tasks - Evaluation of time costs.}

Anomaly detection is formulated	as a binary classification problem in our study. Therefore, we assess the performance of studied models using precision, recall and F1 score. We label the outcomes of these models as true positive (TP), false positive (FP), true negative (TN), and false negative (FN). \accepted{\heng{Explain how they are used to calculate precision, recall and F1 (uncomment some of the text below)?}}
%True positive samples are anomalous log sequences that are correctly classified as anomalies by the models. False positives are normal log sequences that are classified as anomalies. True negatives are normal log sequences that are correctly classified, and false negatives are anomalous log sequences that are falsely considered as normal ones. 
Further, the precision, recall and F1 score are calculated as follows: $Precision = \frac{TP}{TP+FP}$, $Recall = \frac{TP}{TP+FN}$, $F1 = \frac{2PrecisionRecall}{Precision+Recall} = \frac{2TP}{2TP+FP+FN}$. All the metrics are calculated on the test sets.
For some of our results, we only report the $F1$ metric due to space limit. We report the complete results of all the metrics in our replication package.%~\heng{make sure this is done in the replication package}.

%\accepted{\heng{This paragraph is not clear. All the predictions are at the session level, right? Just add one sentence in the previous paragraph should be fine.}}
For sequence-level representations, each sample represents a session, for which classifiers generate one prediction. We calculate the metrics based on predictions for sessions. However, for models that demand fixed-length input, we slice each session with sliding windows and get fixed-length sub-sequences. The labels for these sub-sequences are derived from the session they are from. And models generate predictions for each sliding window. We merge the predictions within sessions and use the labels for sessions to calculate the metrics. 
%\xingfang{rephrased this paragraph.}
%every log sequence is partitioned into several fix-length subsequences for the event-level representations, and models will generate predictions for each sequence. The labels for these sub-sequences are derived from the sequence they are from. So, we can either calculate metrics based on the subsequence or the original sequences by merging the prediction results of the subsequences. To be consistent, we use the latter way to calculate the metrics in all our experiments. \accepted{\Foutse{can you use an example to illustrate these computations clearly?}}

\section{Experimental Results} \label{sec:results}
%\secdescription{Organize your results by RQs. For each RQ, explain your motivation, approach, and results.}

In this section, we present the results of our three research questions, aiming to understand the effectiveness of different log representation techniques in the context of anomaly detection, with the hope that our findings can be generalized to other similar automated log analysis tasks.
%In this section, we organize the quantitative results of the performances of downstream tasks with studied log representations using our three research questions.

\subsection{RQ1. How effective are existing log representation techniques for automated log analysis?} \label{sec:RQ1}

% \xingfang{Rearrange the tables. split one table to two. also add precision and recall.}

\begin{table*}
	\centering
	\vspace{+3mm}
	\caption{Evaluation of six log representation techniques applied to seven anomaly detection models on HDFS dataset.}
	\label{tab:seqlev_hdfs}
    \resizebox{\linewidth}{!}{
    \begin{threeparttable}
            \begin{tabular}{|c|c|l|ccc|ccc|c|} 
\hline
\multicolumn{3}{|c|}{\multirow{2}{*}{Model}}                                                                & \multicolumn{3}{c|}{Classical}                                                                                                                                                                                           & \multicolumn{3}{c|}{Semantic-based}                                         & \multirow{2}{*}{~Gap}  \\ 
\cline{4-9}
\multicolumn{3}{|c|}{}                                                                                      & \multicolumn{1}{c|}{\begin{tabular}[c]{@{}c@{}}Message\\Count\\Vector\end{tabular}} & \multicolumn{1}{l|}{\begin{tabular}[c]{@{}c@{}}TF-IDF\\(ID)\end{tabular}} & \begin{tabular}[c]{@{}c@{}}TF-IDF\\(Text)\end{tabular} & \multicolumn{1}{c|}{~W2V~} & \multicolumn{1}{c|}{FastText} & \multicolumn{1}{c|}{BERT}           &                        \\ 
\hline
\multirow{12}{*}{\rotatebox{90}{Traditional models}} & \multirow{3}{*}{SVM}                                                          & P  & 0.999                                                                               & 0.999                                                                     & 0.999                                                  & 0.998                      & 0.998                         & 0.998          & 0.001                  \\
                       &                                                                               & R  & 0.917                                                                               & 0.999                                                                     & 0.979                                                  & 0.998                      & 0.998                         & 0.998          & 0.082                  \\
                       &                                                                               & F1 & 0.956                                                                               & \textbf{0.999}                                                            & 0.989                                                  & 0.998                      & 0.998                         & 0.998          & 0.043                  \\ 
\cline{2-10}
                       & \multirow{3}{*}{\begin{tabular}[c]{@{}c@{}}Decision\\Tree\end{tabular}}       & P  & 1.000                                                                               & 1.000                                                                     & 0.985                                                  & 0.985                      & 0.985                         & 0.985          & 0.015                  \\
                       &                                                                               & R  & 0.998                                                                               & 0.998                                                                     & 0.999                                                  & 0.998                      & 0.998                         & 0.998          & 0.001                  \\
                       &                                                                               & F1 & \textbf{0.999}                                                                      & \textbf{0.999}                                                            & 0.992                                                  & 0.992                      & 0.992                         & 0.992          & 0.007                  \\ 
\cline{2-10}
                       & \multirow{3}{*}{\begin{tabular}[c]{@{}c@{}}Logistic\\Regression\end{tabular}} & P  & 1.000                                                                               & 0.999                                                                     & 1.000                                                  & 0.999                      & 1.000                         & 0.999          & 0.001                  \\
                       &                                                                               & R  & 0.996                                                                               & 0.997                                                                     & 0.900                                                  & 0.901                      & 0.884                         & 0.999          & 0.115                  \\
                       &                                                                               & F1 & 0.998                                                                               & 0.998                                                                     & 0.947                                                  & 0.948                      & 0.938                         & \textbf{0.999} & 0.061                  \\ 
\cline{2-10}
                       & \multirow{3}{*}{\begin{tabular}[c]{@{}c@{}}Random\\Forest\end{tabular}}       & P  & 0.998                                                                               & 0.999                                                                     & 0.997                                                  & 0.999                      & 0.999                         & 0.998          & 0.002                  \\
                       &                                                                               & R  & 1.000                                                                               & 1.000                                                                     & 0.999                                                  & 0.985                      & 0.985                         & 1.000          & 0.015                  \\
                       &                                                                               & F1 & \textbf{0.999}                                                                      & \textbf{0.999}                                                            & 0.998                                                  & 0.992                      & 0.992                         & \textbf{0.999} & 0.007                  \\ 
\hline
\multirow{9}{*}{\rotatebox{90}{Deep-learning models}}  & \multirow{3}{*}{MLP}                                                          & P  & 0.999                                                                               & 0.911                                                                     & 0.987                                                  & 0.911                      & 0.911                         & 0.911          & 0.088                  \\
                       &                                                                               & R  & 0.999                                                                               & 1.000                                                                     & 0.999                                                  & 0.999                      & 1.000                         & 0.999          & 0.001                  \\
                       &                                                                               & F1 & \textbf{0.999}                                                                      & 0.953                                                                     & 0.993                                                  & 0.953                      & 0.954                         & 0.953          & 0.046                  \\ 
\cline{2-10}
                       & \multirow{3}{*}{CNN}                                                          & P  & \multirow{3}{*}{-}                                                                  & \multirow{3}{*}{-}                                                        & 0.982                                                  & 0.985                      & 0.990                         & 0.992          & 0.010                  \\
                       &                                                                               & R  &                                                                                     &                                                                           & 0.922                                                  & 0.923                      & 0.922                         & 0.921          & 0.002                  \\
                       &                                                                               & F1 &                                                                                     &                                                                           & 0.951                                                  & 0.953                      & \textbf{0.955}                & \textbf{0.955} & 0.004                  \\ 
\cline{2-10}
                       & \multirow{3}{*}{LSTM}                                                         & P  & \multirow{3}{*}{-}                                                                  & \multirow{3}{*}{-}                                                        & 0.991                                                  & 0.997                      & 0.993                         & 0.998          & 0.007                  \\
                       &                                                                               & R  &                                                                                     &                                                                           & 0.922                                                  & 0.921                      & 0.920                         & 0.923          & 0.003                  \\
                       &                                                                               & F1 &                                                                                     &                                                                           & 0.955                                                  & 0.958                      & 0.955                         & \textbf{0.959} & 0.004                  \\
\hline
\end{tabular}
            \begin{tablenotes}
                \small
                % \item[1] The reported metric is F1-Score. Other metrics (precision and recall) are included in our replication package due to space limit.
                \item [1] For each model, the highest F1-Scores achieved by the representation techniques are highlighted.
                \item [2] The 'Gap' columns show the biggest differences between the representation techniques for the dataset.
            \end{tablenotes}
        \end{threeparttable}
}
\end{table*}

% \rotatebox{90}{Deep learning}

\begin{table*}
	\centering
	\caption{Evaluation of six log representation techniques applied to seven anomaly detection models on BGL dataset.
	}
	\label{tab:seqlev_bgl}
	\resizebox{\linewidth}{!}{
    \begin{threeparttable}
        \begin{tabular}{|c|c|l|ccc|ccc|c|} 
\hline
\multicolumn{3}{|c|}{\multirow{2}{*}{Model}}                                                               & \multicolumn{3}{c|}{Classical}                                                                                                                                                                                           & \multicolumn{3}{c|}{Semantic-based}                                         & \multirow{2}{*}{~Gap}  \\ 
\cline{4-9}
\multicolumn{3}{|c|}{}                                                                                     & \multicolumn{1}{c|}{\begin{tabular}[c]{@{}c@{}}Message\\Count\\Vector\end{tabular}} & \multicolumn{1}{c|}{\begin{tabular}[c]{@{}c@{}}TF-IDF\\(ID)\end{tabular}} & \begin{tabular}[c]{@{}c@{}}TF-IDF\\(Text)\end{tabular} & \multicolumn{1}{c|}{~W2V~} & \multicolumn{1}{c|}{FastText} & BERT           &                        \\ 
\hline
\multirow{12}{*}{\rotatebox{90}{Traditional models}} & \multirow{3}{*}{SVM}                                                          & P  & 0.958                                                                               & 0.828                                                                     & 0.855                                                  & 0.853                      & 0.869                         & 0.871          & 0.130                  \\
                      &                                                                               & R  & 0.840                                                                               & 0.654                                                                     & 0.728                                                  & 0.716                      & 0.654                         & 0.667          & 0.186                  \\
                      &                                                                               & F1 & \textbf{0.895}                                                                      & 0.731                                                                     & 0.787                                                  & 0.779                      & 0.746                         & 0.746          & 0.164                  \\ 
\cline{2-10}
                      & \multirow{3}{*}{\begin{tabular}[c]{@{}c@{}}Decision\\Tree\end{tabular}}       & P  & 0.959                                                                               & 0.959                                                                     & 0.971                                                  & 0.781                      & 0.734                         & 0.812          & 0.237                  \\
                      &                                                                               & R  & 0.921                                                                               & 0.919                                                                     & 0.963                                                  & 0.701                      & 0.654                         & 0.701          & 0.309                  \\
                      &                                                                               & F1 & 0.939                                                                               & 0.938                                                                     & \textbf{0.967}                                         & 0.739                      & 0.692                         & 0.752          & 0.275                  \\ 
\cline{2-10}
                      & \multirow{3}{*}{\begin{tabular}[c]{@{}c@{}}Logistic\\Regression\end{tabular}} & P  & 0.947                                                                               & 0.882                                                                     & 0.868                                                  & 0.871                      & 0.844                         & 0.886          & 0.103                  \\
                      &                                                                               & R  & 0.889                                                                               & 0.741                                                                     & 0.728                                                  & 0.753                      & 0.667                         & 0.765          & 0.222                  \\
                      &                                                                               & F1 & \textbf{0.917}                                                                      & 0.805                                                                     & 0.792                                                  & 0.808                      & 0.745                         & 0.821          & 0.172                  \\ 
\cline{2-10}
                      & \multirow{3}{*}{\begin{tabular}[c]{@{}c@{}}Random\\Forest\end{tabular}}       & P  & 0.830                                                                               & 0.810                                                                     & 0.872                                                  & 0.667                      & 0.681                         & 0.694          & 0.205                  \\
                      &                                                                               & R  & 0.963                                                                               & 0.951                                                                     & 0.946                                                  & 0.783                      & 0.808                         & 0.806          & 0.180                  \\
                      &                                                                               & F1 & 0.891                                                                               & 0.875                                                                     & \textbf{0.907}                                         & 0.720                      & 0.738                         & 0.745          & 0.170                  \\ 
\hline
\multirow{9}{*}{\rotatebox{90}{Deep-learning models}}  & \multirow{3}{*}{MLP}                                                          & P  & 0.958                                                                               & 0.951                                                                     & 0.927                                                  & 0.895                      & 0.868                         & 0.910          & 0.090                  \\
                      &                                                                               & R  & 0.840                                                                               & 0.951                                                                     & 0.938                                                  & 0.840                      & 0.815                         & 0.877          & 0.136                  \\
                      &                                                                               & F1 & 0.895                                                                               & \textbf{0.951}                                                            & 0.933                                                  & 0.866                      & 0.841                         & 0.893          & 0.119                  \\ 
\cline{2-10}
                      & \multirow{3}{*}{CNN}                                                          & P  & \multirow{3}{*}{-}                                                                  & \multirow{3}{*}{-}                                                        & 0.900                                                  & 0.868                      & 0.857                         & 0.939          & 0.082                  \\
                      &                                                                               & R  &                                                                                     &                                                                           & 1.000                                                  & 0.975                      & 0.963                         & 0.951          & 0.049                  \\
                      &                                                                               & F1 &                                                                                     &                                                                           & \textbf{0.947}                                         & 0.919                      & 0.907                         & 0.945          & 0.040                  \\ 
\cline{2-10}
                      & \multirow{3}{*}{LSTM}                                                         & P  & \multirow{3}{*}{-}                                                                  & \multirow{3}{*}{-}                                                        & 0.866                                                  & 0.755                      & 0.822                         & 0.871          & 0.116                  \\
                      &                                                                               & R  &                                                                                     &                                                                           & 0.877                                                  & 0.988                      & 0.914                         & 1.000          & 0.123                  \\
                      &                                                                               & F1 &                                                                                     &                                                                           & 0.871                                                  & 0.856                      & 0.865                         & \textbf{0.931} & 0.075                  \\
\hline
\end{tabular}
        \begin{tablenotes}
                \small
                % \item[1] The reported metric is F1-Score. Other metrics (precision and recall) are included in our replication package due to space limit.
                \item [1] For each model, the highest F1-Scores achieved by the representation techniques are highlighted.
                \item [2] The 'Gap' columns show the biggest differences between the representation techniques for the dataset.
            \end{tablenotes}
    \end{threeparttable}
    }
\end{table*}

\begin{table*}
	\centering
	\caption{\response{R1.3 R2.4 R3.1 Updated results}{}Evaluation of six log representation techniques applied to seven anomaly detection models on Spirit dataset.
	}
	\label{tab:seqlev_spirit}
	\resizebox{\linewidth}{!}{
    \begin{threeparttable}
    \begin{tabular}{|c|c|l|ccc|ccc|l|} 
\hline
\multicolumn{3}{|c|}{\multirow{2}{*}{Model}}                                                               & \multicolumn{3}{c|}{Classical}                                                                                                                                                                                           & \multicolumn{3}{c|}{Semantic-based}                                         & \multicolumn{1}{c|}{\multirow{2}{*}{~Gap}}  \\ 
\cline{4-9}
\multicolumn{3}{|c|}{}                                                                                     & \multicolumn{1}{c|}{\begin{tabular}[c]{@{}c@{}}Message\\Count\\Vector\end{tabular}} & \multicolumn{1}{c|}{\begin{tabular}[c]{@{}c@{}}TF-IDF\\(ID)\end{tabular}} & \begin{tabular}[c]{@{}c@{}}TF-IDF\\(Text)\end{tabular} & \multicolumn{1}{c|}{~W2V~} & \multicolumn{1}{c|}{FastText} & BERT           & \multicolumn{1}{c|}{}                       \\ 
\hline
\multirow{12}{*}{\rotatebox{90}{Traditional models}} & \multirow{3}{*}{SVM}                                                          & P  & 0.984                                                                               & 0.978                                                                     & 0.984                                                  & 0.952                      & 0.973                         & 0.973          & \multicolumn{1}{c|}{0.011}                  \\
                      &                                                                               & R  & 0.968                                                                               & 0.963                                                                     & 0.963                                                  & 0.963                      & 0.952                         & 0.968          & 0.016                                       \\
                      &                                                                               & F1 & \textbf{0.976}                                                                      & 0.970                                                                     & 0.973                                                  & 0.957                      & 0.962                         & 0.971          & 0.019                                       \\ 
\cline{2-10}
                      & \multirow{3}{*}{\begin{tabular}[c]{@{}c@{}}Decision\\Tree\end{tabular}}       & P  & 1.000                                                                               & 1.000                                                                     & 1.000                                                  & 0.952                      & 0.962                         & 0.942          & 0.058                                       \\
                      &                                                                               & R  & 0.995                                                                               & 0.995                                                                     & 0.995                                                  & 0.947                      & 0.907                         & 0.952          & 0.088                                       \\
                      &                                                                               & F1 & \textbf{0.997}                                                                      & \textbf{0.997}                                                            & \textbf{0.997}                                         & 0.949                      & 0.934                         & 0.947          & 0.063                                       \\ 
\cline{2-10}
                      & \multirow{3}{*}{\begin{tabular}[c]{@{}c@{}}Logistic\\Regression\end{tabular}} & P  & 0.989                                                                               & 0.989                                                                     & 0.994                                                  & 0.989                      & 0.988                         & 0.984          & 0.010                                       \\
                      &                                                                               & R  & 0.968                                                                               & 0.947                                                                     & 0.925                                                  & 0.947                      & 0.914                         & 0.957          & 0.054                                       \\
                      &                                                                               & F1 & \textbf{0.978}                                                                      & 0.967                                                                     & 0.958                                                  & 0.967                      & 0.950                         & 0.970          & 0.028                                       \\ 
\cline{2-10}
                      & \multirow{3}{*}{\begin{tabular}[c]{@{}c@{}}Random\\Forest\end{tabular}}       & P  & 0.984                                                                               & 0.979                                                                     & 0.982                                                  & 0.941                      & 0.945                         & 0.939          & 0.045                                       \\
                      &                                                                               & R  & 0.984                                                                               & 0.984                                                                     & 0.994                                                  & 0.954                      & 0.957                         & 0.958          & 0.040                                       \\
                      &                                                                               & F1 & 0.984                                                                               & 0.981                                                                     & \textbf{0.988}                                         & 0.947                      & 0.951                         & 0.948          & 0.041                                       \\ 
\hline
\multirow{9}{*}{\rotatebox{90}{Deep-learning models}}  & \multirow{3}{*}{MLP}                                                          & P  & 0.984                                                                               & 0.989                                                                     & 0.989                                                  & 0.978                      & 0.968                         & 0.978          & 0.021                                       \\
                      &                                                                               & R  & 0.957                                                                               & 0.952                                                                     & 0.957                                                  & 0.963                      & 0.968                         & 0.973          & 0.021                                       \\
                      &                                                                               & F1 & 0.970                                                                               & 0.970                                                                     & 0.973                                                  & 0.970                      & 0.968                         & \textbf{0.976} & 0.008                                       \\ 
\cline{2-10}
                      & \multirow{3}{*}{CNN}                                                          & P  & \multirow{3}{*}{-}                                                                  & \multirow{3}{*}{-}                                                        & 0.944                                                  & 0.959                      & 0.935                         & 0.974          & 0.039                                       \\
                      &                                                                               & R  &                                                                                     &                                                                           & 1.000                                                  & 1.000                      & 1.000                         & 1.000          & 0.000                                       \\
                      &                                                                               & F1 &                                                                                     &                                                                           & 0.971                                                  & 0.979                      & 0.966                         & \textbf{0.987} & 0.021                                       \\ 
\cline{2-10}
                      & \multirow{3}{*}{LSTM}                                                         & P  & \multirow{3}{*}{-}                                                                  & \multirow{3}{*}{-}                                                        & 0.940                                                  & 0.943                      & 0.929                         & 0.944          & 0.015                                       \\
                      &                                                                               & R  &                                                                                     &                                                                           & 1.000                                                  & 0.979                      & 0.984                         & 1.000          & 0.021                                       \\
                      &                                                                               & F1 &                                                                                     &                                                                           & 0.969                                                  & 0.961                      & 0.956                         & \textbf{0.971} & 0.015                                       \\
\hline
\end{tabular}
    \begin{tablenotes}
                \small
                \item [1] For each model, the highest F1-Scores achieved by the representation techniques are highlighted.
                \item [2] The 'Gap' columns show the biggest differences between the representation techniques for the dataset.
            \end{tablenotes}
    \end{threeparttable}
    }
\end{table*}

\begin{table*}
	\centering
	\caption{\response{R1.3 R2.4 R3.1 Updated results}{}Evaluation of six log representation techniques applied to seven anomaly detection models on Thunderbird dataset.
	}
	\label{tab:seqlev_thunder}
	\resizebox{\linewidth}{!}{
    \begin{threeparttable}
    \begin{tabular}{|c|c|l|ccc|ccc|c|} 
\hline
\multicolumn{3}{|c|}{\multirow{2}{*}{Model}}                                                               & \multicolumn{3}{c|}{Classical}                                                                                                                                                                                           & \multicolumn{3}{c|}{Semantic-based}                                                   & \multirow{2}{*}{~Gap}  \\ 
\cline{4-9}
\multicolumn{3}{|c|}{}                                                                                     & \multicolumn{1}{c|}{\begin{tabular}[c]{@{}c@{}}Message\\Count\\Vector\end{tabular}} & \multicolumn{1}{c|}{\begin{tabular}[c]{@{}c@{}}TF-IDF\\(ID)\end{tabular}} & \begin{tabular}[c]{@{}c@{}}TF-IDF\\(Text)\end{tabular} & \multicolumn{1}{c|}{~W2V~} & \multicolumn{1}{c|}{FastText} & \multicolumn{1}{c|}{BERT} &                        \\ 
\hline
\multirow{12}{*}{\rotatebox{90}{Traditional models}} & \multirow{3}{*}{SVM}                                                          & P  & 0.999                                                                               & 0.996                                                                     & 0.997                                                  & 0.996                      & 0.992                         & 0.995                    & 0.007                  \\
                      &                                                                               & R  & 1.000                                                                               & 0.999                                                                     & 1.000                                                  & 0.992                      & 0.977                         & 0.983                    & 0.023                  \\
                      &                                                                               & F1 & \textbf{0.999}                                                                      & 0.997                                                                     & 0.998                                                  & 0.993                      & 0.985                         & ~0.989                   & 0.014                  \\ 
\cline{2-10}
                      & \multirow{3}{*}{\begin{tabular}[c]{@{}c@{}}Decision\\Tree\end{tabular}}       & P  & 1.000                                                                               & 1.000                                                                     & 1.000                                                  & 0.985                      & 0.980                         & 0.975                    & 0.025                  \\
                      &                                                                               & R  & 1.000                                                                               & 1.000                                                                     & 1.000                                                  & 0.972                      & 0.952                         & 0.963                    & 0.048                  \\
                      &                                                                               & F1 & \textbf{1.000}                                                                      & \textbf{1.000}                                                            & \textbf{1.000}                                         & 0.979                      & 0.966                         & 0.969                    & 0.034                  \\ 
\cline{2-10}
                      & \multirow{3}{*}{\begin{tabular}[c]{@{}c@{}}Logistic\\Regression\end{tabular}} & P  & 0.999                                                                               & 0.998                                                                     & 0.997                                                  & 0.996                      & 0.996                         & 0.995                    & 0.004                  \\
                      &                                                                               & R  & 0.999                                                                               & 0.981                                                                     & 0.987                                                  & 0.980                      & 0.934                         & 0.977                    & 0.065                  \\
                      &                                                                               & F1 & \textbf{0.999}                                                                      & 0.989                                                                     & 0.992                                                  & 0.988                      & 0.964                         & 0.986                    & 0.035                  \\ 
\cline{2-10}
                      & \multirow{3}{*}{\begin{tabular}[c]{@{}c@{}}Random\\Forest\end{tabular}}       & P  & 0.997                                                                               & 0.999                                                                     & 0.998                                                  & 0.972                      & 0.958                         & 0.966                    & 0.041                  \\
                      &                                                                               & R  & 0.999                                                                               & 0.999                                                                     & 0.999                                                  & 0.993                      & 0.987                         & ~0.994                   & 0.016                  \\
                      &                                                                               & F1 & 0.998                                                                               & \textbf{0.999}                                                            & 0.998                                                  & 0.982                      & 0.972                         & 0.980                    & 0.027                  \\ 
\hline
\multirow{9}{*}{\rotatebox{90}{Deep-learning models}}  & \multirow{3}{*}{MLP}                                                          & P  & 0.998                                                                               & 0.995                                                                     & 0.995                                                  & 0.995                      & 0.972                         & 0.989                    & 0.026                  \\
                      &                                                                               & R  & 0.999                                                                               & 0.997                                                                     & 0.998                                                  & 0.992                      & 0.981                         & 0.992                    & 0.018                  \\
                      &                                                                               & F1 & \textbf{0.999}                                                                      & 0.996                                                                     & 0.997                                                  & 0.993                      & 0.977                         & 0.991                    & 0.022                  \\ 
\cline{2-10}
                      & \multirow{3}{*}{CNN}                                                          & P  &                                                                                     &                                                                           & 0.977                                                  & 0.962                      & 0.955                         & 0.986                    & 0.031                  \\
                      &                                                                               & R  & -                                                                                   & -                                                                         & 1.000                                                  & 1.000                      & 1.000                         & 1.000                    & 0.000                  \\
                      &                                                                               & F1 &                                                                                     &                                                                           & 0.989                                                  & 0.980                      & 0.977                         & \textbf{0.993}           & 0.016                  \\ 
\cline{2-10}
                      & \multirow{3}{*}{LSTM}                                                         & P  &                                                                                     &                                                                           & 0.878                                                  & 0.910                      & 0.875                         & 0.948                    & 0.073                  \\
                      &                                                                               & R  & -                                                                                   & -                                                                         & 1.000                                                  & 1.000                      & 1.000                         & 1.000                    & 0.000                  \\
                      &                                                                               & F1 &                                                                                     &                                                                           & 0.935                                                  & 0.953                      & 0.933                         & \textbf{0.973}           & 0.038                  \\
\hline
\end{tabular}
    \begin{tablenotes}
                \small
                \item [1] For each model, the highest F1-Scores achieved by the representation techniques are highlighted.
                \item [2] The 'Gap' columns show the biggest differences between the representation techniques for the dataset.
            \end{tablenotes}
    \end{threeparttable}
    }
\end{table*}

\subsubsection{Motivation}
Prior works have widely used different log presentation techniques in their automated log analysis workflow. However, no work has comprehensively compared the impact of the choice of log representation techniques in their workflow. 
Therefore, this research question aims to bridge the gap and provide a comprehensive comparison of the commonly used log representation techniques in the context of anomaly detection. %Through the performances of different representations on different follow-up models, we may have a clearer picture of the characteristics and features of existing log representation techniques, thus providing a reference for future designs.
Through analyzing the impact of different log representation techniques on the different anomaly detection models, we hope to provide a reference for future work to choose the appropriate log representation techniques for their specific data, analysis tasks, and use cases. 
%In addition\accepted{\Foutse{why is this contrary?}}, we can also get to know the characteristics \accepted{\Foutse{the argument on intuition is a bit vague!}} of the downstream tasks and follow-up models, which may help us better choose proper input features and a suitable log representation technique for them.

\subsubsection{Approach}

%\heng{As the considered models are common for the three RQs, move them to the Experiment Design section, by adding a small section in ``downstream tasks and datasets''.}
%\xingfang{''downstream tasks and datasets'' exists, I add a subsubsection "Anomaly Detection Models and implementation details" above.}

%We select several representative anomaly detection models to enable a comprehensive and fair comparison of log representation techniques. %For some of them, we make some modifications that will be introduced in the following to adapt them to log representations generated with different log representation techniques.
In this RQ, we evaluate our studied six log representation techniques with seven anomaly detection ML models and four datasets. For each log representation technique, we combine it with each ML model applied to each dataset.

\textit{Combining log representations and ML models.}
As our goal is to evaluate the effectiveness of the existing log representation techniques, we compare the performances of the models of selected downstream tasks with different inputs of representations generated with studied representation techniques.

Message count vector and event template ID-based TF-IDF (TF-IDF ID) are based on the count of log occurrences in log sequences and, thus, can only generate a sequence-level representation for each log sequence. As mentioned before, CNN and LSTM models require event-level log representation due to their mechanisms. Therefore, CNN and LSTM models can not be combined with these two representation techniques. Other representation techniques can generate token-level or event-level log representation. Moreover, low-level log representation can be merged with proper aggregation approaches to higher-level representations. Therefore, these representation techniques can match anomaly detection models that demand both event-level and sequence-level input. For representation techniques that generate token-level representations, we aggregate token-level representation to form event-level representation for a log message. For models requiring sequence-level log representations, we further aggregate the event-level log representations into sequence-level with mean aggregation, which is the most common practice in previous works.

\textit{Using Scott-Knott Effect Size Difference (SK-ESD) test to rank log representation techniques.} To understand the relative rank of the different log representation techniques, we use the SK-EST test~\citep{tantithamthavorn2017mvt, tantithamthavorn2018impact} to rank these techniques into statistically distinct groups based on their performances on studied datasets. We conduct three separate SK-EST tests: One for traditional models, one for deep learning models and a third one for all models.

Different datasets and different downstream models can significantly impact the resulting performance regardless of the chosen log representation techniques. To mitigate such impact when ranking the log representation techniques, we first derive a rank (i.e., initial rank) of each log representation technique for each downstream model applied on each dataset based on the F1 score (i.e., a log representation technique achieving a better F1 score has a higher rank). Each initial rank of a log presentation technique for each model and each dataset serves as one observation for the log representation technique. As we have seven models and four datasets, each log representation technique has 28 observations in total in the overall test. Then, we use the SK-EST test to derive a statistical ranking of the six log representation techniques based on their observations (i.e., initial ranks). The level of significance used in the SK-EST test is set to the default value of 0.05. 

\accepted{\heng{slightly updated}}
\response{R3.4}{As CNN and LSTM models can not be combined with sequence-level representation techniques (i.e., MCV and TF-IDF by message ID), there are some observations\accepted{\heng{observations?}} in the tests (i.e., the MCV and TF-IDF (ID) techniques do not have observations for the CNN and LSTM models).
%has only one observation from each dataset in the deep-only ranking). 
\accepted{\heng{Briefly explain how the missing values are handled}\xingfang{Just left blank for the algorithm.}}
Specifically, the statistical tests underlying the SK-EST tests would be performed with unequal sizes of samples, which may impact the power of the statistical significance~\citep{rusticus2014impact}. Thus, the missing observations may influence the ranking results. We mark the affected representation techniques in Table \ref{tab:sktest}.}

% \xingfang{TODO: deep only/ traditional only ranking, MCV has only one observation for each dataset, which influence the ranking.} 
%We run the test on all studied representations, traditional techniques, and semantic representations, respectively.
%In addition to the overall ranking of all log representation techniques, we also rank their performance for traditional models and deep learning model separately. \heng{check if the above description is correct}

%For each model, we keep hyper-parameters the same for different representations as long as possible to ensure the fairness of the comparisons. In this section, we gives brief introductions to the implementations of adopted downstream models.

\subsubsection{Results}

Table~\ref{tab:seqlev_hdfs},~\ref{tab:seqlev_bgl},~\ref{tab:seqlev_spirit} and~\ref{tab:seqlev_thunder} compare the results of applying different log representation techniques to seven anomaly detection models on the four studied datasets. 
% The results confirm the finding in the previous paper~\citep{he2016experience} that supervised anomaly detection models can generally perform better on the HDFS dataset than on the BGL dataset. The difference is because the HDFS dataset is relatively simple, with only 29 event templates, while BGL has more than 300 templates.
Table~\ref{tab:sktest} shows the statistical rankings of the different log representation techniques from the SK-EST tests.

$\bullet$ \textbf{The choice of log representation techniques has non-negligible influences on the performance of the downstream models.}
As shown in Table~\ref{tab:seqlev_hdfs}, nearly all models achieve very good performance on the HDFS dataset (with F-scores ranging from 0.938 to 0.999), the Spirit dataset (from 0.934 to 0.997), and the thunderbird dataset (from 0.933 to 1.000), while their performance on the BGL dataset is relatively lower (with F-scores ranging from 0.692 to 0.967).
%Our results are similar to that in prior works~\citep{he2016experience,chen2021experience}\heng{is it true?}.
Nevertheless, we observe that different log representation techniques can lead to different performances of the downstream models. On the HDFS dataset, using different log representation techniques causes F-score differences up to 0.061 for the different models; on the BGL dataset, the different log representation techniques lead to F-score differences up to 0.275 for the different models; on the Spirit dataset, the largest discrepancy reaches 0.063, which is 0.038 for the Thunderbird dataset. 

% Some models are more sensitive to the choice to choice of log representation techniques. For example, logistic regression generates F-score differences 0.061, 0.172 respectively for the two datasets, while CNN model only generates 0.004, 0.040, 0.016 and 0.021 differences.
% \heng{give examples based on the Gap value}.

Our SK-EST test results (Table~\ref{tab:sktest}) indicate that there exists statistical difference between the performance of the different log representation techniques. In the overall and traditional-model-only ranking, the six log representation techniques are ranked into five distinct groups. \response{R3.5}{The three classical log representation techniques outperformed their semantic-based counterparts, with MCV achieving the best rank, followed by TF-IDF (ID) and TF-IDF (Text) in the second rank. The BERT embedding is ranked only in the third place, followed by Word2Vec and FastText.
%The variation may come from the fact that different models might highly differ or resemble in their mechanisms. 
%Moreover, 
However, the BERT embedding is ranked first in the deep-model-only ranking, which shows that the deep-learning-based anomaly detection models can generally work better with BERT embedding than traditional models.}

The difference between the performance of different log representation techniques may be explained by the different information represented by different representation techniques. For example, one can directly tell the number of occurrences of certain log events in a sequence from the message count vector. Sometimes, this may be the most critical indicator of an anomaly and lead to a good anomaly detection performance.

%From the results on the HDFS dataset, we find that several anomaly detection models, especially traditional models, achieve near-perfect results, at least on one of the studied representation techniques. Moreover, some models’ performances are stable when matched with all the studied representations. For example, the worst F1 score on HDFS for the random forest classifier is 0.992, which is slightly inferior to the best performance.
%However, performance varies greatly for some models, especially for unsupervised models, when combined with different representation techniques. For example, the largest differences are 0.29 and 0.145 for the isolation forest on two datasets.

\response{R3.4 Begin}{}
\begin{table}[!h]
\centering
\caption{\response{R1.3 R2.4 R3.1 Updated results}{}Statistical ranking of the different log representation techniques from the SK-EST test.} 
\label{tab:sktest}
\resizebox{\linewidth}{!}{
\begin{threeparttable}
% \begin{tabular}{|c|c|c|c|c|} 
% \hline
% \multicolumn{5}{|c|}{Statistically Distinct Groups}                   \\ 
% \hline
% MCV(1) & BERT, ET.TFIDF(2) & TT.TFIDF(3) & Word2Vec(4) & FastText(5)  \\
% \hline
% \end{tabular}
\begin{tabular}{|c|c|c|c|c|c|} 
\hline
\multirow{2}{*}{Model } & \multicolumn{5}{c|}{Statistically Distinct Groups}                                                                                                            \\ 
\cline{2-6}
                        & 1                                                  & 2                                                                   & 3           & 4        & 5         \\ 
\hline
Traditional only        & MCV                                                & \begin{tabular}[c]{@{}c@{}}TF-IDF (Text)\\TF-IDF (ID)\end{tabular}  & BERT        & Word2Vec & FastText  \\ 
\hline
Deep only               & \begin{tabular}[c]{@{}c@{}}MCV$^*$\\BERT\end{tabular} & TF-IDF (Text)                                                       & TF-IDF (ID)$^*$ & Word2Vec & FastText  \\ 
\hline
Overall                 & MCV$^*$                                               & \begin{tabular}[c]{@{}c@{}}TF-IDF (ID)$^*$\\TF-IDF (Text)\end{tabular} & BERT        & Word2Vec & FastText  \\
\hline
\end{tabular}
\begin{tablenotes}
\small
\item[*] The ranking of indicated techniques may be influenced by missing observations\accepted{\heng{observations}}.
% \item[1] Techniques in the same cell are in the same statistically distinct group. For example, TF-IDF (ID) and TF-IDF (Text) are in the same group in the overall ranking.
\end{tablenotes}
\end{threeparttable}
}
\end{table}

\response{R3.4 End}{}

$\bullet$ \textbf{There exists no single log representation technique that performs the best across all models and datasets.}
As shown in Table~\ref{tab:seqlev_hdfs},~\ref{tab:seqlev_bgl},~\ref{tab:seqlev_spirit} and~\ref{tab:seqlev_thunder}, five out of the six log representation techniques (except Word2Vec) achieve the best performance for at least one combination of models and datasets. The best-performing log presentation technique in the overall ranking, Message Count Vector, achieves the best performance for 12 out of the 20 (i.e., 3/5) combinations of models and datasets. However, the technique in the last place (i.e., FastText) achieves the best for only one case out of the 28 combinations.

The findings suggest that researchers and practitioners should be cautious with the selection of log representation techniques and investigate the mechanism of their models and the characteristics of log representation techniques. Based on this knowledge, they can choose the representations that suit their follow-up models best.

\begin{framed}
\noindent\textit{Finding 1:} %Different combinations of log representation techniques and downstream models result in diverse performance.
\accepted{\heng{changed}}The choice of log representation techniques has a non-negligible influence on the performance of the downstream models. While there is no single log representation technique that always performs the best, overall, the simplest message count vector representation performs the best across various models and datasets.
\end{framed}

%\noindent\textbf{For traditional models, classical log representation techniques perform better than semantic-based representation techniques; in comparison, semantic-based representations tend to perform better for deep learning models.}
%\noindent\textbf{For the downstream models that do not consider the ordering of log messages in a session, the classical log representation techniques perform better than semantic-based representation techniques; in comparison, for the downstream models that consider the ordering of log messages in a session, the semantic-based representation techniques perform better.}
%Most traditional models (including all the studied traditional models in this work) and some deep learning models (e.g., MLP) do not consider the ordering of log messages in a session. For these models, event-level log representations are aggregated into the sequence level (i.e., session level). 
%\heng{Xingfang, please confirm if the claims above are right then update the following explanations based on the claims.}\xingfang{not very appropriate. it should have little to do with whether a model takes a sequence or not. mainly because the dimension.}

%For some deep learning models (e.g., CNN and LSTM), event-level log representations are fed directly to the models. Many deep learning models can leverage the sequence of log messages in a session, and thus stand a better chance to perform better.
% \heng{moved this up and highlighted}

$\bullet$ \textbf{Traditional models generally perform better with classical log representation techniques, while deep learning models are able to work well with semantic-based representation.}
For traditional anomaly detection models, 3 out of 4 models achieve the best performance with classical representations on the HDFS dataset and 4 out of 4 on the BGL, Spirit and Thunderbird datasets. %And for unsupervised models, only one best result of the two models on the two datasets is achieved with the semantic representation. 
In total, in 15 out of the 16 cases (four models and four datasets) of traditional machine learning models, classical log representation techniques perform better. The three classical \accepted{\heng{be consistent: use ``classical''}} log representation techniques are listed in the first two statistically distinct groups, which outperformed all their semantic-based counterparts with traditional machine learning models.
\response{R1.3 R2.4 R3.1 R3.5}{However, the results are different for deep anomaly detection models that can leverage the sequential information (i.e., CNN and LSTM)\accepted{\heng{i.e., CNN and LSTM}}: 7 out of 8 cases favour semantic-based embeddings rather than the classical counterpart (i.e., TF-IDF (Text)).\accepted{\heng{compatible classical representation}} But for the MLP, which takes sequence-level\accepted{\heng{sequence-level?}} representation as input, favours (in 3 out of 4 cases) quantitative count-based representations according to our experimental results.}

The performance difference may be caused by the models' discrepancy in the ability to learn the complex and abstract representation of the log data. The classical \accepted{\heng{classical}} representation techniques are based on quantitative or sequential statistics to occurrences of log templates or tokens, whose patterns are relatively simpler than those of semantic embeddings. The semantic-based representations carry higher layer information, which may be utilized by more advanced deep models. \response{R2.2}{The authors of a recent study~\citep{le2021log} utilized a transformer-based model and demonstrated the superiority of semantic embedding over traditional representation. Their comparative experiment indicated that their model performed significantly better with BERT embedding than with the indexes of the log template on some datasets, which confirms our observation}.\accepted{\heng{, which confirms our observation}}

Moreover, event-level log representations are fed directly to the CNN and the LSTM models without the feature aggregation process that transforms event-level features to sequence-level features, which enables them to leverage the sequential information within a log session. This extra information may further boost the performance of these two models.
Another explanation is that the dimensions of traditional representation techniques are usually determined by the vocabulary size or the number of log templates, which are not enormous in some datasets. Traditional models may perform well enough on the low-dimension data. However, deep learning models have more model parameters, which enable the extraction and representation of higher-dimension data, and therefore they can take advantage of the semantic information.

%Some traditional machine learning models perform better on low-dimension data than high-dimension data.

% For some deep learning models (e.g., CNN and LSTM), event-level log representations are fed directly to the models. Many deep learning models can leverage the sequence of log messages in a session, and thus stand a better chance to perform better.

$\bullet$ \textbf{Among the classical log representation techniques, the simplest Message Count Vector technique achieves the best performance; among the semantic-based log representation techniques, the contextual embedding technique BERT achieves the best performance.}
The Message Count Vector technique is the simplest approach and is widely used in automated log analysis tasks~\citep{he2016experience}. Our results show that it achieves the best performance for 12 out of the 20 cases of the models (five models applied to four datasets) that do not leverage the sequential information of log messages. This is also confirmed by the ranking generated by the SK-EST test: In the traditional-only ranking, MCV is ranked in the first place, followed by TF-IDF-based techniques. The BERT is a contextual embedding. It \response{R3.7.}{achieves} the best performance for 7 out of the 8 cases of the models that can leverage sequential information of log (two models applied to four datasets). In the deep-only ranking, BERT is ranked in the first group, which is superior to the other two semantic-based techniques by a large margin. Unlike static embedding, BERT, as a contextual embedding technique, generates representations based on the surrounding context and, thus, is more able to capture the semantic information of a log message. Therefore, representations generated with BERT can achieve good performance with most anomaly detection models.

Therefore, future work can leverage such general rules to choose the appropriate log representation techniques for their models. For traditional models that have limited feature extraction ability, classical representation techniques such as Message Count Vector could be considered. For more sophisticated models with more parameters, semantic-based representation techniques could be considered. And among the semantic-based representations, contextual embeddings may work better than static embeddings.

\paragraph{\responseminor{R2.1}{\textbf{Impact of Different Grouping Settings}}}

\begin{figure}[!ht]
    \centering
	\includegraphics[scale=0.45]{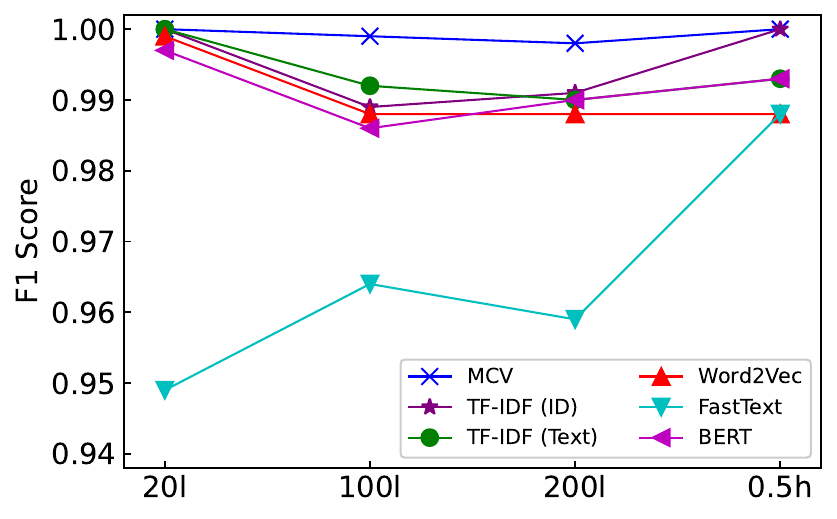}
	\vspace{-3mm}
	\caption[GroupingLength]{\responseminor{R2.1}{Results of logistic regression model with different grouping settings on Thunderbird dataset.}
	}
	\vspace{-2mm}
	\label{fig:g_length}
\end{figure}

\pararesponseminor{Impact of Different Grouping Settings. Different log sequence lengths resulting from different grouping settings could impact the representations and the performance of models, which was shown in experiments from previous works (e.g., RQ2 in \citet{le2022log}). When grouping the studied datasets, we adopted different grouping settings, hoping that our findings could be tenable across varying settings. In particular, we follow prior works using the same datasets to config the group settings. To further examine the impacts that variations of the grouping process may have, we conduct an additional evaluation, in which we group the Thunderbird dataset with different fixed window settings (i.e., 20 logs, 100 logs, 200 logs, and 0.5-hour logs), which are in accordance with the settings in \citet{le2022log}. Figure \ref{fig:g_length} shows the results.}
% Table \ref{tab:g_length} shows the results.}

% \usepackage{multirow}

% \begin{table}[h!]
% \centering
% \label{tab:g_length}
% \caption{\responseminor{R2.1}{Results of logistic regression model with different grouping settings on Thunderbird dataset}}
% \resizebox{.5\linewidth}{!}{
% \begin{tabular}{ccccc} 
% \hline
% \multirow{2}{*}{Techniques} & \multicolumn{4}{c}{Grouping Window Size}  \\ 
% \cline{2-5}
%                             & 20l   & 100l  & 200l  & 0.5h              \\ 
% \hline
% MCV                         & 1.000 & 0.999 & 0.998 & 1.000             \\
% TF-IDF (ID)                 & 1.000 & 0.989 & 0.991 & 1.000             \\
% TF-IDF (Text)               & 1.000 & 0.992 & 0.990 & 0.993             \\
% Word2Vec                    & 0.999 & 0.988 & 0.988 & 0.988             \\
% FastText                    & 0.949 & 0.964 & 0.959 & 0.988             \\
% BERT                        & 0.997 & 0.986 & 0.990 & 0.993             \\
% \hline
% \end{tabular}
% }
% \end{table}

\pararesponseminor{Based on the obtained results, it is evident that performance variations can occur due to different grouping configurations. In general, the relative ranking among the log representation techniques remains the same for the different lengths of log sequences. There may be multiple factors contributing to these performance variations, making it challenging to accurately evaluate the effectiveness of log representation techniques across various grouping settings. For example, discrepancies in dataset size can impact performance variations since the composition and size of the training and test sets are influenced by different grouping configurations. Furthermore, there is no best-performing setting for all the studied techniques according to the results. As a result, we believe that the lengths of log sequences intricately influence both log representations and models through complex mechanisms. Future evaluations should focus on investigating the effects of grouping settings on log representation techniques.
}

\begin{framed}
%\noindent\textit{Finding 2: } 
%Traditional anomaly detection models favour numerical representations generated by traditional techniques, while deep learning models generally favour semantic embeddings.
%\heng{this finding can be made more interesting if we show that prior studies use traditional models with semantic embeddings, as well as using deep models with tradition techniques. Some examples can be found in papers ``Can Pre-trained Code Embeddings Improve Model Performance? Revisiting the Use of Code Embeddings in Software Engineering Tasks'' and ``Using Black-Box Performance Models to Detect Performance Regressions under Varying Workloads: An Empirical Study'', though anomaly detection papers can be more relevant}

%Downstream models that do not consider log ordering in a session (including almost all traditional models) favor classical log representation techniques, with the Message Count Vector technique performing the best; in comparison, downstream models that consider log ordering in a session (including many deep learning models such as CNN and LSTM) favor the semantic-based representation techniques, with the contextual embedding technique BERT performing the best.

\noindent\textit{Finding 2:} 
Traditional anomaly detection models perform well on classical log representations. However, deep models can achieve better performance with semantic-based representations by their stronger feature extraction and representation ability; Among the classical log representation techniques, Message Count Vector \response{R3.7.}{achieves} the best performance. Context embedding (BERT) generally performs better among the semantic-based log representation techniques.
\end{framed}

\subsection{RQ2. How does log parsing influence the effectiveness of log representations in automated
log analysis?} \label{sec:RQ2}

\subsubsection{Motivation}
The log parsing process transforms semi-structured raw logs into structured data by separating variables from log messages and retaining the log templates. Log parsing is a common pre-processing step before the log representation step. 
%In automated log analysis, log parsing is sometimes an imperative step for feature extraction and representation to separate variable parts from log messages and retain log templates. 
%The log parsing process can reduce the noises in the log data, thus lowering the analyzing difficulty of follow-up models.
Although many log parsers with different mechanisms have been developed and achieved high performance and high accuracy~\citep{zhu2019tools}, the errors introduced by the parsing process may sometimes undermine the performances of log analysis according to the empirical study of Le et al.~\citep{le2021log}. Essential words may be removed from a parsing error which results in information loss. As log parsing may be error-prone and cause information loss, some researchers have explored some log analysis frameworks~\citep{le2021log} that take raw logs as inputs. 
It is not clear how log parsing and log representation together impact the performance of downstream tasks. 
%, may generate inaccurate results. Some automated log analysis frameworks abandon the log parser and take raw logs as inputs. 
Thus, in this RQ, we investigate the potential impacts that log parsing, when used with different log representation techniques, may have on the performance of downstream models.

%In this research question, we want to quantify the impacts of the log parsing process has on the performances of downstream tasks. Whether the log parser can cause any information loss or eliminate the noises from the log data is a question we need to answer when looking into the representation approaches. With the investigation, we may get a fairer conclusion about the effectiveness of representations. Moreover, further investigating the consequences of log parser may help researchers design new representation approaches.

\subsubsection{Approach}
%\secdescription{Describe the specific approach for this RQ. }

In this RQ, we consider the log representation techniques that are compatible with both parsed and unparsed log data. Then we compare the performance of the downstream models that take the representations built from parsed and unparsed log data. 

% For investigating the impact of log parsing on representations, we generated log representation for both original log messages and parsed logs with representation techniques. We evaluate the quality of these representations by the performances of models that are based on them.

\textit{Selection of log representation techniques.}
From the studied log representation techniques, we select Word2Vec and FastText to answer this research question. The representation generated by these two techniques can remain the same regardless of the configuration of log parsing, which is not the case for representation like Log Template Text-based TF-IDF (TF-IDF Text), whose dimension may vary according to the vocabulary of the corpus in the dataset. As the dimension can also impact on the performance of models, we choose techniques that can generate fixed dimension representations for both parsed and unparsed log data. Also, the high dimension and enormous model size of pre-trained BERT prohibits us to generate features for unparsed logs in our server.

% It integrates vector generation for some out-of-vocabulary(OOV) words by separating them into subwords. In original log messages, there are many tokens for which other techniques can not generate embedding. By utilizing FastText, we can reduce the impact of OOVs.
% The output dimension of the pre-trained FastText models can be adjusted easily, which enable us to generate feature vectors for datasets and get evaluation results with our limited computation and storage resources, as sometimes feature vectors generated for datasets may be extremely large.

\textit{Comparison of using parsed and unparsed log data to build representations.}
We compare and analyze the performances of the studied anomaly detection models with the features generated by these two representation techniques with both parsed and unparsed logs. Moreover, we also generate the visualization of embeddings with a dimension reduction algorithm (t-SNE~\citep{van2008visualizing}) to get some intuitions from the data to better explain the varied results.

\subsubsection{Results}

\begin{figure}[!t] \centering
	\includegraphics[scale=0.35]{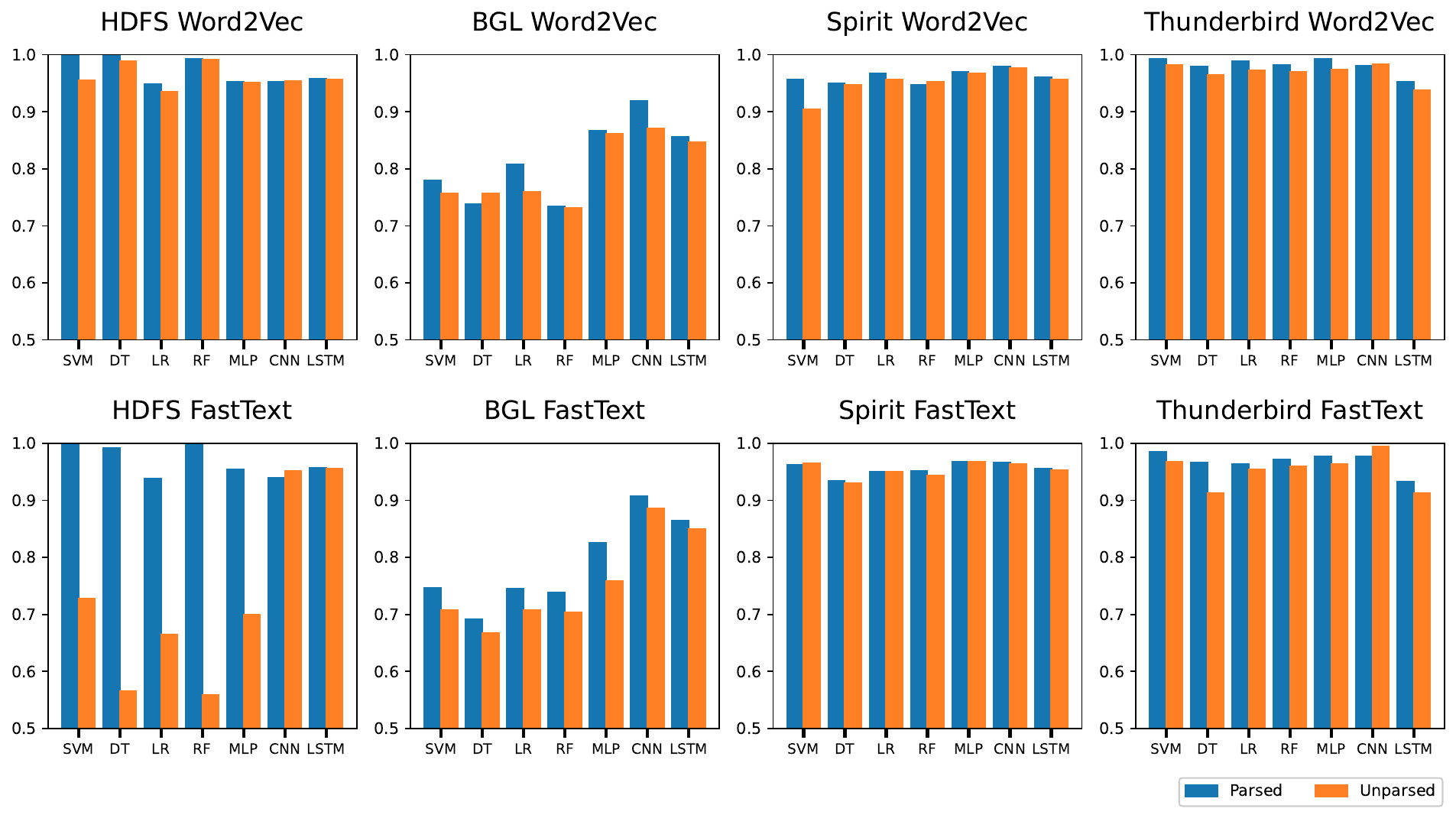}

	\caption[parse]{\response{R1.3 R2.4 R3.1 (added results for two additional datasets.)}{}Comparison of performances of the studied anomaly detection models using the Word2Vec and FastText representations that are generated from parsed and unparsed logs.
	%\heng{Only reports F1-score to keep consistency with other results; use two back to back bars like in RQ3 for parsed vs. unparsed logs}
	}
	\label{fig:parse}

\end{figure}

% \heng{this part goes to the approach section (selection of log representation techniques)}
%The goal of log parsing is to distinguish between constants and variables in log messages and generate log templates. If there is no parsing process, some traditional log representation technique may not works. For example, the log text-based TF-IDF presentation will generate an extremely high-dimensional representation for unparsed log sequences if the dynamic areas in log messages are not properly removed. Each new token will take one more dimension in the final representation. Let alone the message count representation, which directly requires log templates ID as input. This research question evaluates with representations generated with pre-trained FastText model.

Fig.\ref{fig:parse} shows the comparison of performances of studied models with FastText representations generated with original and parsed log messages. 

$\bullet$  \textbf{In general, log parsing improves the quality of the generated log representations and thereby the performance of the downstream models.} For the HDFS dataset, the two log representation techniques, Word2Vec and FastText, achieve average performance (F1-score) improvements of 0.010 and 0.236 across the seven models, respectively. For the BGL dataset, the average improvements are 0.017 and 0.034. \response{R1.3 R2.4 R3.1}{For Spirit, the improvements are 0.010 and 0.002, which are 0.012 and 0.016 for Thunderbird.}

% For the HDFS dataset, the two log representation techniques, Word2Vec and FastText, achieve performance (F1-score) improvements of 0 to 0.043 and 0.002 to 0.44 across the seven models, respectively. For the BGL dataset, the improvements are 0 to 0.44 and 0.016 to 0.066. For Spirit, the improvements are -0.005 to 0.053, -0.003 to 0.007. For Thunderbird, the improvements are -0.003 to 0.016 and -0.018 to 0.053.
%From the bar chart, we find that most of the studied anomaly detection models perform poorly with representations of unparsed logs. However, we find that the performances are similar between representations generated with parsed and unparsed logs for CNN and LSTM models. 

\begin{figure}[!t]
	\centering 
	\subfigure[Unparsed Logs]{  
		\begin{minipage}{0.38\textwidth}
			\centering 
			\includegraphics[width=4.5cm]{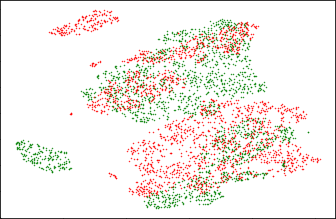}
			\label{fig:viscontent}
		\end{minipage}
	}
	\subfigure[Parsed Logs]{ 
		\begin{minipage}{0.38\textwidth}
			\centering 
			\includegraphics[width=4.5cm]{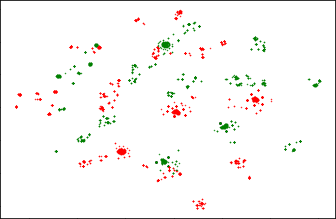}
			\label{fig:vistemplate}
		\end{minipage}
	}
	\caption[Visualization of log representations with FastText]{Visualization of representations generated with FastText using t-SNE. 200 positive (red) and negative (green) samples are randomly sampled from the HDFS dataset.}
	\label{fig:visemb}
\end{figure}

%And for the unsupervised methods, PCA performs similarly with these two inputs. 
In particular, for the HDFS dataset with the FastText representation, parsing leads to a very large difference in the performance of five out of the seven studied models. We randomly sample the FastText representations of 200 positive and negative samples from the HDFS dataset and use t-SNE~\citep{van2008visualizing} to visualize them. The visualization in Figure~\ref{fig:visemb} shows that representations for parsed logs are more compact than those of unparsed logs, which means the embeddings generated with the parsed logs are more distinguishable than those generated from the original log messages. However, deep learning models may work better with unparsed log data in some cases. For example, on HDFS and Thunderbird datasets, CNN performs better with unparsed logs by a small margin. The reason behind this may be that the parsing errors induced by the log parser can undermine the performance. The impact of log parsing errors was also examined in~\citet{le2021log}'s work.

% \heng{Update the following text based on new results. Re-organize the text to support the two bold sentences.}

The characteristic of the representation technique can explain the general inferior performances on unparsed logs: There is no proper mechanism to represent numerical values or special tokens in logs for these representation techniques. The representations generated for these tokens would be a noise in feature representation if they are not treated as OOVs. 

$\bullet$ \textbf{Depending on the datasets, some models (e.g., CNN and LSTM) are less sensitive to whether the log data is parsed or not.} CNN and LSTM perform similarly with the two different inputs may be a little counter-intuitive. One possible explanation is that these two deep sequential models have strong feature extraction and representation abilities and can offset the impacts of the noise. At the same time, unparsed logs will not introduce noises caused by the parsing errors.

% Also, unlike other models which take sequence-level representation as input, these two models take event-level log representation as input. The feature aggregation process that transforms representation from event-level to sequence-level may cause information loss.

Although there exist log analysis frameworks that take unparsed logs as input, to our best knowledge, they adopt a preprocessing process to manually remove parameters or other fields from raw logs~\citep{le2021log}, which can be regarded as a 'vanilla' parsing process. \response{R2.7}{If certain fields in logs, such as numerical values, special tokens, and error codes, are not adequately preprocessed, modelled, and utilized, it may have an adverse effect on the representation of the log.} This finding implies that careful preprocessing and modelling of these fields are crucial for optimal log representation. Log parsing is an effective way to remove these unrecognizable texts for pre-trained language models and thus reduce the noise in representations. Future researchers and practitioners should pay attention to the preprocessing process before adopting log representation techniques, even if they abandon the log parsing process when designing a log analysis framework. \response{R2.7}{Additionally, to improve the overall performance, future researchers and practitioners may also want to take into account the modelling of certain fields (e.g., component name, CPU usage, the time elapsed for a certain process, etc.)\accepted{\heng{give a simple example}} that cannot be embedded by language models but are critical to their downstream tasks~\citep{du2017dl}.}

\paragraph{\response{R3.6}{\textbf{Impact of Refining the Parsing Results}}}
\pararesponse{
As mentioned in \ref{sec:data_prep}, we utilized additional regular expressions to improve the parsing results. We then did a sensitivity test to see its potential impact on the performance. We further evaluated our previously studied representation techniques using the Thunderbird dataset parsed by the Drain parser with regular expressions and trained logistic regression models. The results are shown in Table \ref{tab:regexp}.
}
\begin{table}[!h]
\centering
\label{tab:regexp}
\caption{\response{R3.6}{Performance of logistic regression model on Thunderbird dataset parsed without extra regular expressions. The values under the F1 Scores indicate differences compared with corresponding results in RQ1.}} 
\resizebox{.8\linewidth}{!}{
\begin{tabular}{ccccccc} 
\hline
Technique & \begin{tabular}[c]{@{}c@{}}Message\\Count\\Vector\end{tabular} & \begin{tabular}[c]{@{}c@{}}TF-IDF\\(ID)\end{tabular} & \begin{tabular}[c]{@{}c@{}}TF-IDF\\(Text)\end{tabular} & W2V                                                    & FastText                                           & BERT                                           \\ 
\hline
F1 Score  & \begin{tabular}[c]{@{}c@{}}0.999\\(=)\end{tabular}             & \begin{tabular}[c]{@{}c@{}}0.989\\(=)\end{tabular}   & \begin{tabular}[c]{@{}c@{}}0.991\\(0.001$\downarrow$)\end{tabular} & \begin{tabular}[c]{@{}c@{}}0.987\\(0.001$\downarrow$)\end{tabular} & \begin{tabular}[c]{@{}c@{}}0.964\\(=)\end{tabular} & \begin{tabular}[c]{@{}c@{}}0.983\\(0.003$\downarrow$)\end{tabular}  \\
\hline
\end{tabular}
}
\end{table}

\pararesponse{
Although we can tell that the parsing results are refined by the observation that repetitive templates are decreased, we only observed minor accuracy gains for some representations after passing the regular expressions from the experiment. In contrast to the previous findings, which demonstrated that parsed and unparsed logs could lead to significant discrepancies, the refinement of parsing outcomes did not have a substantial impact on performance. This could be attributed to the ability of machine learning models to learn how to exclude unimportant features or irrelevant noise.
}

\pararesponse{However, a large number of error templates may greatly increase the dimension of some representation techniques (e.g., for MCV, the dimension is equal to the number of resulting templates.). A large number of error templates increases the learning burden when we train the follow-up models. Sometimes it may even make the model training unprocurable.}

\paragraph{\responseminor{R2.2}{\textbf{Impact of Using Different Log Parsers}}}
\pararesponseminor{Recent studies~\citep{dai2020logram, khan2022guidelines, liu2022uniparser} adopt new metrics to evaluate the existing log parsers. Apart from just reporting the Group Accuracy of the parsing results, these works report other metrics (e.g., Parsing Accuracy, Edit Distance, etc.), which may give a more comprehensive evaluation of a parser. While it is true that the Drain parser achieves a high group accuracy, it presents inferior results in some metrics (i.e., Parsing Accuracy) in some recent works.}

\pararesponseminor{A higher Group Accuracy may benefit the representation techniques that rely on log templates (e.g., MCV). In contrast, a high Message-Level accuracy may contribute to the quality of representations based on the token-level census or embedding generation (e.g., TF-IDF (Text)). Therefore, we conducted another sensitivity test to reduce our evaluation’s potential bias. In this test, we adopt the LogPPT parser~\citep{le2023log}, which exhibits superior results over different metrics, including Parsing Accuracy, to further evaluate the quality of the studied representation techniques using the Thunderbird dataset and trained logistic regression models. The results are shown in Table \ref{tab:logppt}.}

\begin{table}[!h]
\centering
\label{tab:logppt}
\caption{\responseminor{R2.2}{Performance of logistic regression model on Thunderbird dataset parsed by LogPPT parser. The values under the F1 Scores indicate differences compared with corresponding results in RQ1.}} 
\resizebox{.8\linewidth}{!}{
\begin{tabular}{ccccccc} 
\hline
Technique & \begin{tabular}[c]{@{}c@{}}Message\\Count\\Vector\end{tabular} & \begin{tabular}[c]{@{}c@{}}TF-IDF\\(ID)\end{tabular}              & \begin{tabular}[c]{@{}c@{}}TF-IDF\\(Text)\end{tabular} & W2V                                                                & FastText                                                            & BERT                                                                \\ 
\hline
F1 Score  & \begin{tabular}[c]{@{}c@{}}0.999\\(=)\end{tabular}             & \begin{tabular}[c]{@{}c@{}}0.996\\(0.007 $\uparrow$)\end{tabular} & \begin{tabular}[c]{@{}c@{}}0.992\\(=)\end{tabular}     & \begin{tabular}[c]{@{}c@{}}0.981\\(0.007$\downarrow$)\end{tabular} & \begin{tabular}[c]{@{}c@{}}0.941\\(0.023 $\downarrow$)\end{tabular} & \begin{tabular}[c]{@{}c@{}}0.985\\(0.001$\downarrow$)\end{tabular}  \\
\hline
\end{tabular}
}
\end{table}

\pararesponseminor{From the results of this sensitive test, it is evident that there exist slight variations in performance when parsing the dataset using a different parser. The exploration of correlations or patterns between the performances of log parsers and the quality of log representation techniques is yet to be conducted. This presents an avenue for future evaluations in the realm of log parsers, representation techniques and downstream models.}

\begin{framed}
\noindent \textit{Finding 3:} 
%Log parsing is sometimes an essential step for some log representation techniques and follow-up learning processes. 
In general, log parsing improves the quality of the generated log representations and, thereby, the performance of the anomaly detection models. It reduces the noise in representations and thus alleviates models' learning burden by removing dynamic fields in logs. \response{R2.7}{Proper preprocessing and modelling of these dynamic\accepted{\heng{recognizable by whom? language models?}} fields may be crucial for optimal log representation.}
\accepted{\heng{I feel this is not supported by our results, and may be removed from the summary box, what do you think?}\xingfang{I think it is fine, just a small extension to the experimental results.}}
\end{framed}

\subsection{RQ3. How do representation aggregation methods influence the effectiveness of log representation in automated log analysis?} \label{sec:RQ3}

\subsubsection{Motivation}

% Different log representation techniques can generate the representation at different levels (e.g., token level, log event level, or sequence level).

% Depending on the requirements of the downstream models, low-level representations need to be aggregated into high-level ones according to the need of the follow-up models. 

% There exist different aggregation methods (e.g., mean aggregation). However, it is not clear how different aggregation methods impact the performance of the different representation techniques no the downstream models. In this RQ, we aim to explore the potential influence of different aggregation methods when used together with different log representation techniques and different models.

%For this research question, we will investigate the feature aggregation in log representation from the following two perspectives.

%First, 
For representation techniques that generate word embeddings(e.g., Word2Vec, FastText) for tokens in log events, we need to merge these token-level representations into event-level ones. The related works usually used mean aggregation to form the representation for log events~\citep{meng2019loganomaly}, in which information may be lost, as some keywords that carry essential semantic information and severity in logs may be diluted by averaging. Therefore, we aim to compare different aggregation methods and evaluate the impacts they have on the quality of log representation.

%Second, most representation techniques can generate event-level representations for log data apart from some traditional representation approaches like message count vector and log key. However, different follow-up models require different levels of input features. So, we sometimes need to merge features of a lower level to a higher level in order to accommodate the model. There are different approaches to merge the features. In this research question, we focus on evaluating the impacts that different feature aggregation methods have on the performance of downstream tasks. We compare the most common feature aggregation approaches from automated log analysis models.

%Second, 
In addition, for sequential models that take a fixed length of log messages as input, the event-level representation will be implicitly aggregated by the models to generate analytical results according to its task. So, we need to partition the session with a fixed-length window and a pre-defined step size. This implicit aggregation may also influence the performances of downstream models. Therefore, we investigate the impact that different configurations of session partition may have on the performance of downstream tasks. Although results generated with window-based inputs will be merged to generate the final predictions for log sessions, we want to quantify the impacts of different configurations of aggregation on the downstream tasks.

\subsubsection{Approach}
% For the evaluation of feature aggregation approaches (i.e., from token-level to event-level, event-level to sequence-level), we select the two most common ones, the mean aggregation and the TF-IDF
% aggregation.

%\heng{In this research question, we investigate the impact of feature aggregation in log representation from two perspectives: 1) method of aggregating token-level representations, and 2) aggregation window size of sequential models.} \xingfang{accepted}
In this research question, we investigate the impact of feature aggregation in log representation from two perspectives: 1) the method of aggregating token-level representations, and 2) the aggregation window size of sequential models.

\emph{Method of aggregating token-level representations.}
For the first perspective, we select the two most common aggregation practices, the \textbf{mean aggregation} and the \textbf{TF-IDF aggregation}~\citep{chen2021experience}.
% \heng{references} \heng{also need to explain how you do mean aggregation even though it might be straightforward} 
For mean average aggregation, we aggregate the token-level representations by averaging the feature vectors by each dimension. For TF-IDF aggregation, we calculate the TF-IDF values for each token in log templates and calculate the weighted average of the token-level representation to form the event-level representation for log events. Moreover, we use the mean average to aggregate them into sequence-level representation. We select Word2Vec and FastText as they generate token-level representations.

\emph{Window size for feature aggregation in sequential models.}
For the second perspective, we study the implicit aggregation process within the sequential models. From Table~\ref{tab:Sliding_params}, we can find the significant difference in the average size of sessions in the four studied datasets. So, we conducted some preliminary experiments to broadly define the suitable range of the window size and further pre-defined some specific window sizes accordingly to investigate the impacts of implicit aggregation of studied sequential models. \response{R1.3 R2.4 R3.1}{The chosen window size range for the HDFS and the Thunderbird is between 10 to 50, while for the BGL and the Spirit, whose sessions are usually longer, the range is 20 to 80.} We adopt all studied techniques that can generate event-level representations~(i.e., except the Message Count Vector and TF-IDF (ID) which can only generate sequence-level representations).

% As the average number of log events in a session varies with the datasets, we select different ranges and step sizes accordingly\heng{this sentence belongs to the approach}. \xingfang{duplicated information}

% \heng{provide details about the chosen window sizes}  \heng{provide details of ``all the techniques''}

% In this RQ, we apply different aggregation methods on the resulting representations from different representation techniques and evaluate the performance on the downstream models. We study the aggregation process that merges feature from token-level to event-level as it is applicable to all our studied models. 

%However, for the second perspective, we calculate the TF-IDF values for each template, weigh the event-level embedding, and merge them to form the sequence-level representation.

% To investigate the impacts of implicit aggregation of sequential models, we pre-defined some window sizes according to features of the studied datasets (i.e., the ranges of the numbers of log events that a log session usually contains.)
 
\subsubsection{Results}

% \xingfang{No. of Anomalies / No. of total logs}

\begin{figure}[!t] \centering
	\includegraphics[scale=0.35]{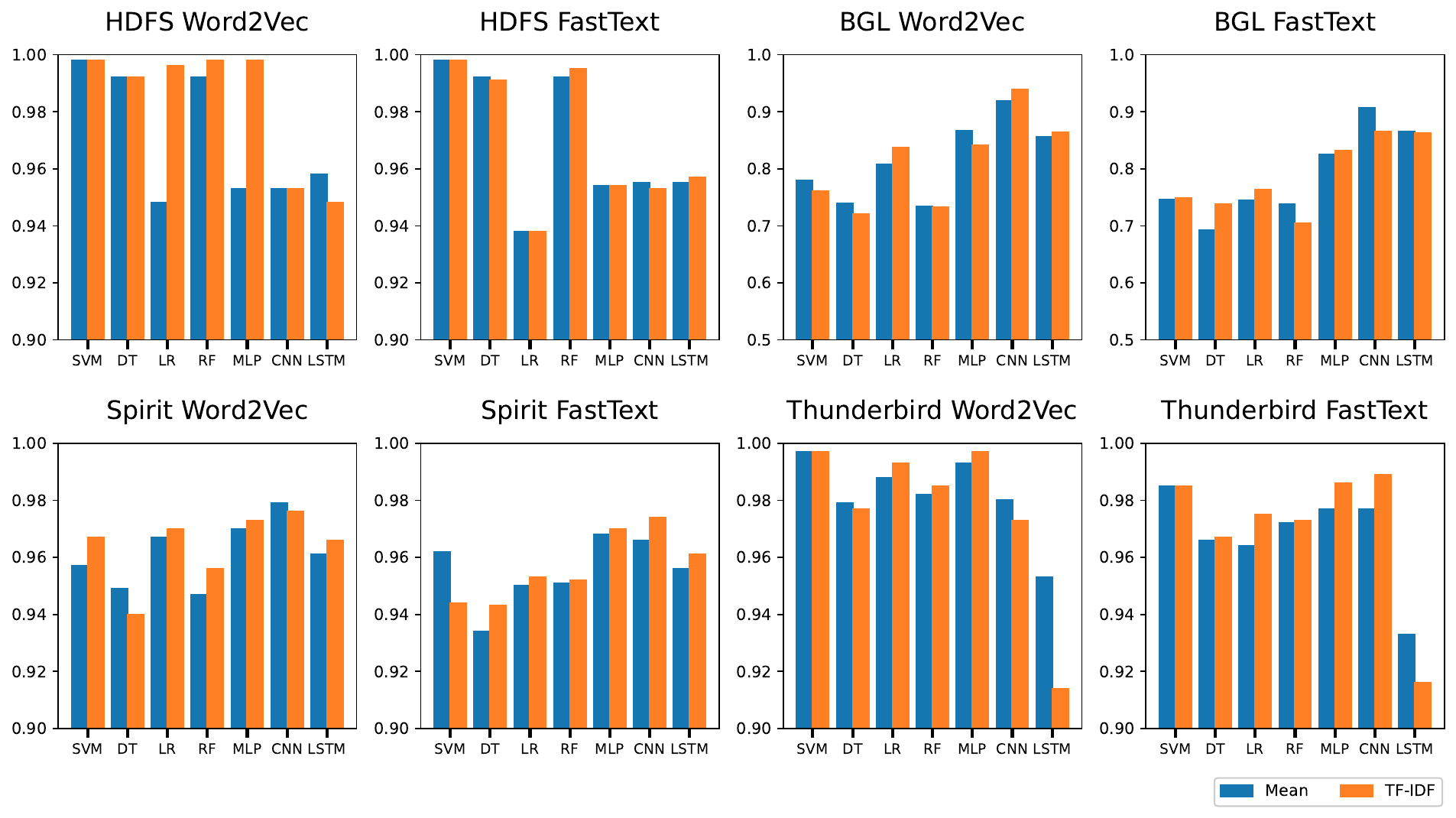}
	\caption[aggregation]{\response{R1.3 R2.4 R3.1 Updated results (added results for two additional datasets.)\accepted{\heng{be more specific, like: Updated results (added results for two additional datasets). Also applies to other figures/tables}}}{}Comparison of performances of FastText Log representation with TF-IDF and Mean aggregation with LSTM anomaly detection model.}
	\label{fig:agg}
\end{figure}

$\bullet$ \textbf{Different approaches of aggregating representations can cause non-negligible differences in the performance of the downstream models.}
From figure \ref{fig:agg}, we can tell that there exist some performance gaps between these two aggregation methods on some combinations of the dataset, model and log representation. For example, the logistic regression model may favour TF-IDF aggregation with both representation techniques on all studied datasets: TF-IDF aggregation outperformed the mean aggregation in all eight cases. However, this conclusion is invalid on other representation techniques and follow-up models. This finding indicates that the aggregation approaches can have non-negligible impacts on the effectiveness of log representation and thus influence the performance of log analysis models.

%Based on this finding, future works can try different approaches of feature aggregation to seek the best performance of their log analysis models.

% Apart from the window size, merging low-level log representations to high-level ones is another factor that will influence the feature aggregation and, thus, impact the quality of the representations. Here, we use the feature embedding generated with a pre-trained FastText model with the LSTM anomaly detection model on the HDFS dataset to evaluate the mean average and TF-IDF aggregation. The aggregation here refers to the process which merges embedding for tokens in a log message to the log-level representation.

$\bullet$ \textbf{However, there is no clear pattern on which aggregation method performs better, as the impacts to the performance of aggregation method vary according to the combination of dataset, model, and representation techniques.} 
From the results, we can not find a clue to tell which aggregation method works better: For Word2Vec representation on the HDFS dataset, 3 out of 7 models perform significantly better with TF-IDF aggregation. However, this is not the case for other datasets and representation combinations. Different combinations of the dataset, model and representation favour different aggregation methods. Moreover, the difference in performance also varies among the combinations. Some combinations may be more sensitive to the utilization of aggregation techniques.

% \begin{framed}
% \noindent \textit{Finding 4: } %Different feature aggregation approaches may influence the quality of the feature.
% Different aggregation methods can cause non-negligible difference in the performance of the follow-up models, while there is no clear pattern on which aggregation method perform generally better. 
%Researchers and developers should try different approaches of feature aggregation to seek the best performances.
% \end{framed}

\begin{figure}[!t] \centering
	\includegraphics[scale=0.35]{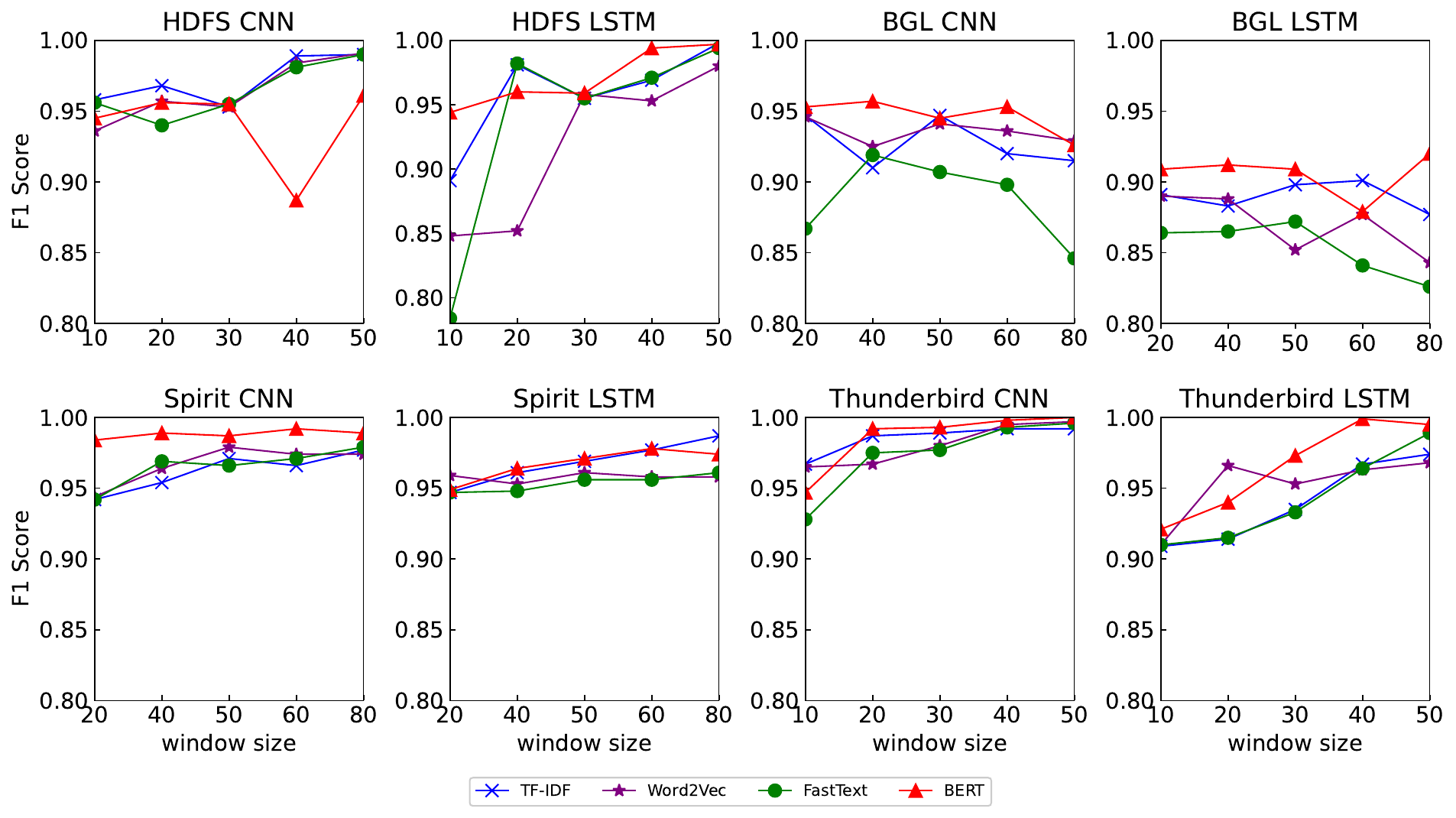}
	\caption[win size]{\response{R1.3 R2.4 R3.1 Updated results}{}The impacts of different window sizes for feature aggregation.}
	\label{fig:winsize}
\end{figure}

% \xingfang{window-size analysis added. The following paragraphs are new.}

$\bullet$ \textbf{Window size for sequential models can significantly affect the performances of downstream models.}
\response{R1.3 R2.4 R3.1 Updated results}{}
Figure~\ref{fig:winsize} shows the F1 scores of CNN and LSTM models with different representations varying according to the input window size on four studied datasets. From the graph~\ref{fig:winsize}, we can see the fluctuation of the F1 score with the variation in window size. On the HDFS dataset, the biggest difference in the F1 score for the CNN model is 0.074, achieved by BERT, which is 0.210, achieved by FastText for the LSTM model. On the BGL dataset, the biggest difference in F1 score is 0.073, achieved by FastText and 0.047, achieved by Word2Vec. \response{R1.3, R2.4, R3.1}{On the Spirit dataset, the biggest difference in the F1 score for the CNN model is 0.037, achieved by FastText, and 0.04 for the LSTM model, achieved by TF-IDF. On the Thunderbird dataset, the biggest gaps for CNN and LSTM are 0.068 and 0.079, respectively, both achieved by FastText.} The results show that the window size for feature aggregation can pose nonnegligible impacts on the performance of anomaly detection task.

% Generally, the performance improves as the window size increases for the HDFS dataset. However, the performance slightly drops after the window size of 50 on the BGL dataset. 

% \heng{the results are not discussed to support the message! Give some concrete numbers, like up to how much (percentage or difference range) difference} \xingfang{explaination added.}

$\bullet$ \textbf{The differences in performance may be caused by the intrinsic features of datasets.}
The line charts show an improvement in performance when window size increases for the HDFS on almost all the studied log representation techniques with two models. We do not expand the range’s upper bound for HDFS as the F1 score almost reaches 1, and the window size of 50 is larger than the length of most sessions in the dataset. For the BGL dataset, the peak is generally around 50, and the performances tend to decrease thereafter. \response{R1.3, R2.4, R3.1}{For Spirit and Thunderbird, we also observed growth in performance when increasing the window size, while the variations are more stable compared with the other\accepted{\heng{other}} two datasets.}

% \heng{people will ask whether it is because you choose different window size ranges for the two datasets: for HDFS the maximum is 50 while for BGL it peaks at 50. Need to either use a similar range (increase the range of HDFS to also show a peak) or provide a very strong justification why choosing different ranges.} \xingfang{fixed}
\accepted{\heng{Highlight this, one key information is that: For the same dataset, window size affects the different representation techniques in a similar way. remember the RQ title is about how aggregation methods impact log representation (don't move too far away from log representation in the discussions).}}

$\bullet$ \textbf{For the same dataset, window size affects the different representation techniques in a similar way.} The variation trends for different combinations of log representations and models are generally consistent on the same dataset, with some outliers (e.g., BERT with CNN when window size is 40, Word2Vec with LSTM when the window size is 20). Therefore, we believe that the intrinsic features of the dataset cause the differences in performance.
% However, the partition configurations generally have similar impacts on different log representations, although there exist some discrepancies. 

\response{R1.1 R3.7 Begin}{}

\pararesponse{More specifically, the characteristics of anomalies in a log sequence vary according to datasets. The lengths of abnormal sequences may have different ranges in different datasets. The sliding window setting can influence the distribution of anomalies in models’ input windows, and some continuous anomaly log sequences may be truncated into multiple input windows\accepted{\heng{into multiple input windows}} in some input windows. Therefore, the sliding window setting may have a significant\accepted{\heng{significant}} impact on the performance. Similarly to this, the aforementioned grouping methods, which group a log sequence into sessions,\accepted{\heng{not clear what are grouping methods}} can also have impacts on the performance of different anomaly detection models, which were found by recent work~\citep{le2022log}. In their work, their finding suggests that the performances of models suffer when dealing with shorter log sequences.}

\pararesponse{
The impacts of sliding window settings may\accepted{\heng{maybe add a ``may'' here, as the connection from the previous graph is not quite supported}} vary mainly depending on the datasets\accepted{\heng{used datasets?}}. 
% follow-up models, log representation techniques, and even grouping techniques.
\accepted{\heng{I would suggest only focusing on depending on datasets (you have support from Fig 6: the trends are generally consistent for each dataset. We don't have support for other aspects}}It is a great challenge for the developer to determine the most suitable sliding window setting for their cases, as it may demand onerous experiments. \response{R2.9}{Besides, we notice that recent studies introduce Graph Neural Networks (GNN)~\citep{xie2022loggd, wan2021glad} to log representation, and the experiments from these works show that these models are robust against the variation of window size. Future researchers may utilize more stable log representation techniques, which are less sensitive to the variation of feature aggregation settings, to ensure more stable performances of their models.}}

\response{R1.1 R3.7 End}{}
% \heng{First mention that the impact is significant}
\begin{framed}
%It is worth experimenting on different window sizes for feature aggregation with target data sources to achieve better performances.
% \noindent \textit{Finding 4:} Different aggregation configurations can cause non-negligible differences in the performance of the follow-up models, while there is no clear pattern on which aggregation settings may generally perform better. \response{R3.7}{The impact of aggregation settings may vary on the used dataset, follow-up models, log representation techniques, and grouping techniques\accepted{\heng{this sentence is not related to aggregation. maybe say: The impact of aggregation settings depends on the used datasets, ...?}} \accepted{\heng{I would suggest focusing on dataset: For the same dataset, window size affects the different representation techniques in a similar way. The characteristics of the datasets determine the impact of the aggregation window...}\xingfang{We merged the finding 4 and 5, this finding should also cover the results of using tfidf and mean aggregation}}. There is no single best configuration for all. Future researchers can either consider this factor by experimenting with different feature aggregation settings or utilizing log representation techniques that are more stable to different aggregation settings.\accepted{\heng{can we recommend the stable representations based on our results?}\xingfang{It is hard to tell which techniques are more stable from our experiments.}}}

\noindent \textit{Finding 4:} Different aggregation configurations can cause non-negligible differences in the performance of the follow-up models, while there is no clear pattern on which aggregation settings may generally perform better. \response{R3.7}{The different impact of the aggregation configurations on the downstream model performance may be caused by the intrinsic features of datasets. In particular, for the same dataset, the window size affects the different representation techniques in a similar way. Future researchers and practitioners are suggested to explore different feature aggregation settings by considering the characteristics of the datasets or utilizing log representation techniques that are more stable to different aggregation settings.}

\end{framed}

\begin{comment}
\begin{figure}[H] \centering
	\includegraphics[scale=0.4]{./fig/tfidf_winsize.pdf}
	\caption[aggregation]{Performance of studied representations on different window sizes for implicit aggregation of sequential models.}
	\label{fig:winsize}
\end{figure}

\begin{figure}[!htbp]
	\centering 
	\subfigure[HDFS CNN]{  
		\begin{minipage}{0.21\textwidth}
			\centering 
			\includegraphics[width=3.8cm]{./fig/HDFS_CNN.pdf}
		\end{minipage}
	}
	\subfigure[HDFS LSTM]{ 
		\begin{minipage}{0.21\textwidth}
			\centering 
			\includegraphics[width=3.8cm]{./fig/HDFS_LSTM.pdf}
		\end{minipage}
	}
	\subfigure[BGL CNN]{  
		\begin{minipage}{0.21\textwidth}
			\centering 
			\includegraphics[width=3.8cm]{./fig/BGL_CNN.pdf}
		\end{minipage}
	}
	\subfigure[BGL LSTM]{ 
		\begin{minipage}{0.21\textwidth}
			\centering 
			\includegraphics[width=3.8cm]{./fig/BGL_LSTM.pdf}
		\end{minipage}
	}
	\caption[win size]{Window size.}
	\label{fig:winsize}
\end{figure}
\end{comment}

\section{Discussions}
\label{sec:discussions}

% \secdescription{Discuss important take-home messages or cross-cutting findings that combine different RQs.}

%\begin{comment}
%\subsection{Findings}
%\heng{this subsection may be commented out for now, the most important implications can be briefly discussed in the conclusions section.}
In this section, based on our results for answering our research questions, we discuss the implications of our findings. Additionally, we summarize the key factors to consider when selecting the most appropriate log representation techniques for log-based anomaly detection approaches or other automated log analysis tasks. Our recommendations and discoveries will be helpful to researchers and practitioners in selecting the optimal log representation techniques and achieving favourable outcomes in their automated log analysis frameworks.
% We summarize the characteristics of the studied log representation techniques to assist with this process.

\subsection{Implications}

$\bullet$ \textbf{Automated log analysis approaches should pay attention to the choice of log representation techniques as they have a non-negligible impact on the follow-up models.}
Existing log-based anomaly detection approaches usually consider only a single log representation technique. For example, in the work that adopts CNN to detect anomalies in log sequences~\citep{lu2018detecting}, only log keys are used to learn the embeddings for log events, and information from log parameters and messages is lost in this process. 
% In DeepLog~\citep{du2017dl}, although parameters in log events are considered together with the log key, there still exists a possibility that the performance can be improved with other approaches that include semantic information of log messages. 
Our results suggest that the performance of these approaches may be improved by considering other ways of log representations. Also, new representation approaches may be developed according to the specific tasks and downstream models. In particular, classic machine learning models may favour representations generated by traditional techniques. In contrast, deep-learning-based downstream models can better utilize semantic embedding to achieve better results. Also, experiments show that contextual embedding performs the best among the pre-trained language models. Our findings can provide guidance for future work to choose and design the appropriate log representation techniques for their specific tasks. For example, researchers should consider the capability of feature extraction and representation of the models they adopt when choosing the log representation techniques. Models of higher complexity (with more parameters) are more capable of dealing with higher dimensional representations.

% we observe that the simplest message count vector representation works well with classic machine learning models, while high-dimension semantic embedding can achieve better results when matched with deep learning models. 

$\bullet$ \textbf{Log parsing or other preprocessing are recommended before the log representation process as they usually improve the performance of the downstream log analysis tasks.}
% or at least does not harm
Most of the prior works on log-based automated log analysis adopt a log parser to transform raw logs into structured data. Recent work~\citep{le2021log} explores omitting the parsing process and extracting and representing information directly from the raw log data. However, they usually employ some preprocessing steps to remove unnecessary fields in log data.
We find that the log parsing process generally positively impacts automated log analysis, although sometimes it may be time-consuming and erroneous. Also, log parsing enables template-based log representation techniques and removes dynamic fields that will hinder the other semantic-based techniques.
Thus, we suggest that researchers should carefully consider whether to employ the log parsing process in their workflow. As log parsers may sometimes be error-prone and consume additional computational resources, researchers can consider substituting them with lightweight preprocessing processes.

$\bullet$ \textbf{Log analysis workflows should consider experimenting with different configurations of feature aggregation.}
When aggregating low-level log representations to high-level ones, prior works (e.g.,~\citep{chen2021experience}) usually adopt a single strategy without experimenting with other configurations. 
However, according to our experiments, the feature aggregation process is essential for log representation. Feature aggregation configurations can significantly impact downstream models’ performances. However, the impacts are closely related to the characteristics of the datasets. Prior works stand a good chance of achieving better performances when employing different feature aggregation configurations. Therefore, we advise researchers to consider the intrinsic features of the studied log data and employ different configurations when designing their automated log analysis workflow.

\response{R1.1 R3.7 Begin}{}
\subsection{Key factors for selecting log representation techniques}

\pararesponse{To provide insights for researchers and practitioners in selecting appropriate log representation techniques, we summarize below the key factors that need to be considered based on our experiments and findings. We recommend researchers and practitioners consider these factors in their log-based anomaly detection and potentially other automated log analysis tasks to achieve optimal performance in such tasks.} 

\pararesponse{$\bullet$ \textbf{Quality of representation} The quality of log representations is a crucial factor that significantly affects the performance of downstream models. In our study, we found that different models can benefit from different log representations. Across various models and datasets, we determined that the simplest message count vector representation can perform well in most cases. In addition, traditional anomaly detection models generally performed well on classical log representations, while deep models achieved better performance with semantic-based representations due to their stronger feature extraction and representation ability. Among the classical log representation techniques, the Message Count Vector approach achieved the best performance, while Context embedding (BERT) generally performed better among the semantic-based log representation techniques. Therefore, selecting high-quality log representation techniques is essential for achieving optimal downstream model performance.}

\pararesponse{$\bullet$ \textbf{Dimension of representation}
One of the key factors to consider when selecting representation techniques is the dimension of the resulting representation. For some representation techniques, their resulting dimensions are data-invariant, which means the dimensions will remain the same when they are applied to different log data. Semantic-based techniques (e.g., BERT) and graph-based techniques that utilize a neural network structure to generate embeddings for log data can usually provide fixed-length outputs. By contrast, classical techniques (e.g., message count vector) usually rely on a vocabulary of tokens or log templates and thus, the dimensions are subjective to the data. The advantages of techniques with fixed output dimensions are obvious: First, they can better serve the scenarios when data shifting (e.g., vocabulary changes) exists in system logs caused by software evolution. When new log templates appear, these techniques are able to encode new templates while maintaining the feature property. Second, fixed output dimensions may enable more stable performances over different datasets on the same model. When working with datasets with a larger number of log templates, the anomaly detection models may suffer from a performance loss due to their limited model capacity. Higher dimension input usually demands a larger model with more parameters. Classical techniques may generate representations of a wide range of dimensions over different datasets. For example, MCV generates vectors of 46 dimensions on the HDFS dataset, while for the Thunderbird, the dimension is 1,488 in our experiments. Higher dimensions may lead to higher computational costs in the training and prediction stages of follow-up downstream models.}

\pararesponse{$\bullet$ \textbf{Need for log parsing}
As we discussed previously in RQ2, the log parsing process can generally remove noises caused by unprocessed tokens in log data, while errors induced by log parsers may cause performance loss~\citep{le2022log}. Besides, the log parsing process can be time-consuming and require significant manual and computational resources. While log parsing is not essential for semantic-based representation techniques since they typically do not require log template information to operate, it may still be included as a preprocessing step for the logs. In this situation, a complete parsing process may be substituted by a lightweight preprocess, in which tokens that can not be processed by vectorizers are removed, when getting log templates is not mandatory for the representation technique.}

\pararesponse{$\bullet$ \textbf{Computational cost for representation construction}
Another important consideration is the computational cost. Besides the log parsing process, log representation techniques require computational resources (time and space) to convert raw logs, log templates, or log template IDs to numeric vectors. As the mechanism varies across different techniques, the differences in computational cost are significant. For classical techniques, much memory may be used to construct dictionaries and vectorize tokens varying with datasets. For semantic-based techniques, although programmers can utilize the off-the-shelf pre-trained models to escape the computational consumption for training the language models, some techniques still require heavy computations to acquire embeddings. For example, contextual embedding techniques require more computational resources to construct representations than static embedding techniques. Pre-trained BERT models process input tokens through transformer blocks, which involve significant computation and sometimes require specialized hardware (e.g., GPUs, TPUs), particularly for lengthy texts. By comparison, Word2Vec uses a shallow neural network, which is computationally efficient, to generate word embeddings. Taking this factor into account is important when designing anomaly detection workflows that are targeted for online or real-time application scenarios.}

\pararesponse{$\bullet$ \textbf{Granularity}
It is mandatory to ensure that the level of log representation is aligned with the specific anomaly detection model being used. This is because different models require varying levels of granularity and types of information from the log data to perform according to their varying mechanisms. Semantic-based representation techniques (e.g., Word2Vec) can usually generate token-level representations, which can be aggregated into higher-level ones, while some classical techniques can only work on higher-level representations (e.g., MCV can only generate sequence-level representation). Therefore, it is crucial to carefully consider the requirements of the anomaly detection model being employed and choose the log representation accordingly.}

\pararesponse{$\bullet$ \textbf{Explainability}
Finally, explainability is another factor to consider when selecting a log representation technique. Usually, classical log representation techniques (e.g., MCV), which represent the quantitative characteristics of log sequences, have better explainability compared with their semantic-based counterparts, which are learning-based. With a good explainability of log representation techniques, researchers can better understand the prediction given by the follow-up models and, therefore, are able to trace the roots when performance is not satisfactory. Poor explainability of log representation techniques will make the decision-making process a black box, in which the decision-making process becomes agnostic. Future researchers should consider this factor when designing a trustworthy automated log analysis system.}

\pararesponse{In conclusion, selecting an appropriate log representation technique requires careful consideration of several factors, including the quality of representation, dimension of representation, need for log parsing, computational cost for representation construction, granularity, and explainability.}

\response{R1.1 R3.7 End}{}

\section{Threat to Validity}
\label{sec:threats}

We have identified the following threats to the validity of our findings:

%\noindent $\bullet$
\noindent \textbf{External validity.} We carried out this research only based on the log anomaly detection task with the hope that our experimental results and findings can serve as a reference for other automated log analysis tasks. The conclusions may not apply to other downstream tasks, as different downstream tasks or models may have different intrinsic characteristics and favour different configurations or features of log representation. \response{R3.2}{However, anomaly detection is one of the most studied downstream tasks in the domain of automated log analysis~\citep{fu2009execution, xu2009detecting, he2016experience, chen2021experience, meng2019loganomaly, le2021log, wang2018anomaly, zhang2019robust, du2017dl, lu2018detecting, nedelkoski2020self}, demonstrating its importance and popularity. Due to the fact that many automated log analysis tasks share similar
% , if not the same \heng{don't mention ``the same'' as they are never the same}, 
pipelines\accepted{\heng{pipelines?}} that process log data, our work may\accepted{\heng{use ``may'' as we did not prove it}} also inspire and support the designs of workflows of other tasks, despite the fact that only log-based anomaly detection is studied in our work.}

\response{R2.2}{As the mechanism of anomaly detection approaches differs greatly, we limit our research to the supervised log-based anomaly detection models to ensure a fair comparison among studied representations. Therefore, our findings may not apply to unsupervised methods. Future work that examines the impact of log representations on unsupervised learning tasks can complement our results.} \accepted{\heng{Future work that examines the impact of log representations on unsupervised learning tasks can complement our results.}}

\response{R2.2 R2.9}{Recently, new approaches (e.g., Transformer-based~\citep{nedelkoski2020self, le2021log}, graph-based approaches~\citep{xie2022loggd, wan2021glad}) have been applied to log-based anomaly detection.
% \heng{cut off the rest of the sentence: duplicate}by some recent works. 
However, we did not evaluate them in this work, as the mechanisms of these approaches differ greatly, which makes it hard for us to fit them into our research questions. Future works may\accepted{\heng{use ``could'' or ``may'' to tone down a little}} further examine these new approaches and their susceptibility to log representation techniques. To compensate for this, we discussed\accepted{\heng{discussed}} the most representative transformer-based approaches, and related\accepted{\heng{related}} the findings from these works with our experimental results and findings.}

%\noindent $\bullet$ 
\noindent \textbf{Construct validity.} We followed some existing works in the experiment to use pre-trained models trained with natural language models. The experimental results may not reflect the true capability of these log representation techniques, as the effectiveness of generated log representations may suffer greatly from OOV issues or incorrect semantics caused by the different characteristics between log data and natural language.

%\noindent $\bullet$ 
\noindent \textbf{Internal validity.} Our configurations for dataset partition may not be optimal and may influence the accuracy of the evaluation. According to our survey, different log anomaly detection works adopt different grouping configurations for the studied public datasets. We referred to previous works and chose the most common grouping configurations to enable a better comparison. \response{R1.3 R2.4 R3.1}{In addition\accepted{\heng{In addition}}, we employed different grouping configurations for the four studied datasets with the hope that our results and findings can be invariant to different grouping settings of datasets.} Further study may\accepted{\heng{may or could}} be carried out to evaluate the impacts of the data grouping on the log representations. \response{R2.3}{The hyperparameters for the machine learning models in our studies might not be fully optimized. Instead of aiming for the best performance for each particular model, our main goal was to examine how well alternative log representation strategies performed across various\accepted{\heng{various or different}} downstream models. As a result, we made sure that each representation technique was applied to the same dataset with identical parameter settings. Besides,\accepted{\heng{Besides,}} our experimental results are generally consistent with those of prior studies that employed similar datasets, representations, and models. Additionally, we have included our implementations in our replication package, making it possible to reproduce our results. These factors help to mitigate the potential impact of \accepted{\heng{using}}using suboptimal hyperparameters in our study.}\response{R2.2}{Instead of directly assessing the quality of representations, we rely on the performance of downstream models as an indirect measure. However, the variables involved in these downstream tasks may affect the internal validity and introduce potential biases. To mitigate the potential bias, we consider multiple datasets and downstream models in our experiments.}\response{R2.5}{In addition, while most models performed well on the four datasets we examined, we found that the choice of log representation technique could affect downstream model performance. We did observe differences in F-scores when using different log representation techniques. These variations were statistically significant as confirmed by our SK-EST analysis.}\response{R3.4}{Furthermore, certain techniques' characteristics in our SK-Test resulted in some missing observations that could affect the ranking of the studied representation techniques, which we have indicated by clearly indicating the affected techniques.}

\section{Conclusions} \label{sec:conclusions}

%\secdescription{Briefly summarize the approach and results. Then highlight the take-home messages and opportunities for future work.}

%Logs are an essential source of information that software practitioners can observe to understand the runtime states of software systems. The log analysis tasks are gradually automated thanks to various log analysis approaches, in which log representation techniques play an important role. However, researchers or developers are not fully aware of the characteristics of various log representation techniques. They may not make the right decisions when designing their automated log analysis approaches due to the lack of comparison among current log representation techniques. 
% \heng{Remove the text above as they are not conclusions.}
Our work makes a comprehensive evaluation and review of six log representations on four public datasets with seven supervised anomaly detection models. We also examine the impacts of log parsing and feature aggregation of features on the effectiveness and quality of log representations. Our findings suggest that log representation techniques can significantly impact the performance of downstream models. We provide some general guidance and key factors in choosing suitable representation techniques. Also, we find that log parsing can generally improve the quality of log representation by reducing noise in some representation techniques. Moreover, the impacts of configuration for feature aggregation may vary according to the representation, data and downstream models. When designing an automated log analysis workflow, these factors should be carefully taken into account by researchers and engineers. For future work, we plan to evaluate log representation techniques with more automated log analysis downstream tasks and try to explore different features that different downstream tasks may favour. Thus, we can provide a more comprehensive direction for researchers to design their automated log analysis frameworks. 

\section*{Conflicts of Interests}
The authors have no competing interests to declare that are relevant to the content of this article.

\bibliographystyle{natbib}
\bibliography{main}   % name your BibTeX data base

\begin{thebibliography}{}

\bibitem[Chen {\em et~al.}(2004)Chen, Zheng, Lloyd, Jordan, and
  Brewer]{chen2004failure}
Chen, M., Zheng, A.~X., Lloyd, J., Jordan, M.~I., and Brewer, E. (2004).
\newblock Failure diagnosis using decision trees.
\newblock In {\em International Conference on Autonomic Computing, 2004.
  Proceedings.}, pages 36--43. IEEE.

\bibitem[Chen {\em et~al.}(2021)Chen, Liu, Gu, Su, and Lyu]{chen2021experience}
Chen, Z., Liu, J., Gu, W., Su, Y., and Lyu, M.~R. (2021).
\newblock Experience report: Deep learning-based system log analysis for
  anomaly detection.
\newblock {\em arXiv preprint arXiv:2107.05908\/}.

\bibitem[Chow {\em et~al.}(2014)Chow, Meisner, Flinn, Peek, and
  Wenisch]{chow2014mystery}
Chow, M., Meisner, D., Flinn, J., Peek, D., and Wenisch, T.~F. (2014).
\newblock The mystery machine: End-to-end performance analysis of large-scale
  internet services.
\newblock In {\em 11th USENIX Symposium on Operating Systems Design and
  Implementation (OSDI 14)\/}, pages 217--231.

\bibitem[Dai {\em et~al.}(2022)Dai, Li, Chen, Shang, and Chen]{dai2020logram}
Dai, H., Li, H., Chen, C.-S., Shang, W., and Chen, T.-H. (2022).
\newblock Logram: Efficient log parsing using $n$n-gram dictionaries.
\newblock {\em IEEE Transactions on Software Engineering\/}, {\bf 48}(3),
  879--892.

\bibitem[Devlin {\em et~al.}(2018)Devlin, Chang, Lee, and
  Toutanova]{devlin2018bert}
Devlin, J., Chang, M.-W., Lee, K., and Toutanova, K. (2018).
\newblock Bert: Pre-training of deep bidirectional transformers for language
  understanding.
\newblock {\em arXiv preprint arXiv:1810.04805\/}.

\bibitem[Du {\em et~al.}(2017)Du, Li, Zheng, and Srikumar]{du2017dl}
Du, M., Li, F., Zheng, G., and Srikumar, V. (2017).
\newblock Deeplog: Anomaly detection and diagnosis from system logs through
  deep learning.
\newblock In {\em Proceedings of the 2017 ACM SIGSAC conference on computer and
  communications security\/}, pages 1285--1298.

\bibitem[El-Sayed {\em et~al.}(2017)El-Sayed, Zhu, and
  Schroeder]{el2017learning}
El-Sayed, N., Zhu, H., and Schroeder, B. (2017).
\newblock Learning from failure across multiple clusters: A trace-driven
  approach to understanding, predicting, and mitigating job terminations.
\newblock In {\em 2017 IEEE 37th International Conference on Distributed
  Computing Systems (ICDCS)\/}, pages 1333--1344. IEEE.

\bibitem[Fu {\em et~al.}(2009)Fu, Lou, Wang, and Li]{fu2009execution}
Fu, Q., Lou, J.-G., Wang, Y., and Li, J. (2009).
\newblock Execution anomaly detection in distributed systems through
  unstructured log analysis.
\newblock In {\em 2009 ninth IEEE international conference on data mining\/},
  pages 149--158. IEEE.

\bibitem[Fu {\em et~al.}(2013)Fu, Lou, Lin, Ding, Zhang, and
  Xie]{fu2013contextual}
Fu, Q., Lou, J.-G., Lin, Q., Ding, R., Zhang, D., and Xie, T. (2013).
\newblock Contextual analysis of program logs for understanding system
  behaviors.
\newblock In {\em 2013 10th Working Conference on Mining Software Repositories
  (MSR)\/}, pages 397--400. IEEE.

\bibitem[Grave {\em et~al.}(2018)Grave, Bojanowski, Gupta, Joulin, and
  Mikolov]{grave2018learning}
Grave, E., Bojanowski, P., Gupta, P., Joulin, A., and Mikolov, T. (2018).
\newblock Learning word vectors for 157 languages.
\newblock In {\em Proceedings of the International Conference on Language
  Resources and Evaluation (LREC 2018)\/}.

\bibitem[Hansen and Atkins(1993)Hansen and Atkins]{hansen1993automated}
Hansen, S.~E. and Atkins, E.~T. (1993).
\newblock Automated system monitoring and notification with swatch.
\newblock In {\em LISA\/}, volume~93, pages 145--152. Monterey, CA.

\bibitem[He {\em et~al.}(2016a)He, Zhu, He, Li, and Lyu]{he2016evaluation}
He, P., Zhu, J., He, S., Li, J., and Lyu, M.~R. (2016a).
\newblock An evaluation study on log parsing and its use in log mining.
\newblock In {\em 2016 46th annual IEEE/IFIP international conference on
  dependable systems and networks (DSN)\/}, pages 654--661. IEEE.

\bibitem[He {\em et~al.}(2017)He, Zhu, Zheng, and Lyu]{he2017drain}
He, P., Zhu, J., Zheng, Z., and Lyu, M.~R. (2017).
\newblock Drain: An online log parsing approach with fixed depth tree.
\newblock In {\em 2017 IEEE international conference on web services (ICWS)\/},
  pages 33--40. IEEE.

\bibitem[He {\em et~al.}(2016b)He, Zhu, He, and Lyu]{he2016experience}
He, S., Zhu, J., He, P., and Lyu, M.~R. (2016b).
\newblock Experience report: System log analysis for anomaly detection.
\newblock In {\em 2016 IEEE 27th international symposium on software
  reliability engineering (ISSRE)\/}, pages 207--218. IEEE.

\bibitem[He {\em et~al.}(2020)He, Zhu, He, and Lyu]{he2020loghub}
He, S., Zhu, J., He, P., and Lyu, M.~R. (2020).
\newblock Loghub: a large collection of system log datasets towards automated
  log analytics.
\newblock {\em arXiv preprint arXiv:2008.06448\/}.

\bibitem[He {\em et~al.}(2021)He, He, Chen, Yang, Su, and Lyu]{he2021survey}
He, S., He, P., Chen, Z., Yang, T., Su, Y., and Lyu, M.~R. (2021).
\newblock A survey on automated log analysis for reliability engineering.
\newblock {\em ACM Computing Surveys (CSUR)\/}, {\bf 54}(6), 1--37.

\bibitem[Jarry {\em et~al.}(2021)Jarry, Kobayashi, and
  Fukuda]{jarry2021quantitative}
Jarry, R., Kobayashi, S., and Fukuda, K. (2021).
\newblock A quantitative causal analysis for network log data.
\newblock In {\em 2021 IEEE 45th Annual Computers, Software, and Applications
  Conference (COMPSAC)\/}, pages 1437--1442. IEEE.

\bibitem[Katkar and Kasliwal(2014)Katkar and Kasliwal]{katkar2014use}
Katkar, D. G.~S. and Kasliwal, A.~D. (2014).
\newblock Use of log data for predictive analytics through data mining.
\newblock {\em Current Trends In Technology And Science\/}, {\bf 3}(3).

\bibitem[Khan {\em et~al.}(2022)Khan, Shin, Bianculli, and
  Briand]{khan2022guidelines}
Khan, Z.~A., Shin, D., Bianculli, D., and Briand, L. (2022).
\newblock Guidelines for assessing the accuracy of log message template
  identification techniques.
\newblock In {\em Proceedings of the 44th International Conference on Software
  Engineering\/}, pages 1095--1106.

\bibitem[Le and Zhang(2021)Le and Zhang]{le2021log}
Le, V.-H. and Zhang, H. (2021).
\newblock Log-based anomaly detection without log parsing.
\newblock In {\em 2021 36th IEEE/ACM International Conference on Automated
  Software Engineering (ASE)\/}, pages 492--504. IEEE.

\bibitem[Le and Zhang(2022)Le and Zhang]{le2022log}
Le, V.-H. and Zhang, H. (2022).
\newblock Log-based anomaly detection with deep learning: how far are we?
\newblock In {\em 2022 IEEE/ACM 44th International Conference on Software
  Engineering (ICSE)\/}, pages 1356--1367. IEEE.

\bibitem[Le and Zhang(2023)Le and Zhang]{le2023log}
Le, V.-H. and Zhang, H. (2023).
\newblock Log parsing with prompt-based few-shot learning.
\newblock {\em arXiv preprint arXiv:2302.07435\/}.

\bibitem[Li {\em et~al.}(2020)Li, Chen, Jing, He, and Yu]{li2020swisslog}
Li, X., Chen, P., Jing, L., He, Z., and Yu, G. (2020).
\newblock Swisslog: Robust and unified deep learning based log anomaly
  detection for diverse faults.
\newblock In {\em 2020 IEEE 31st International Symposium on Software
  Reliability Engineering (ISSRE)\/}, pages 92--103. IEEE.

\bibitem[Liang {\em et~al.}(2007)Liang, Zhang, Xiong, and
  Sahoo]{liang2007failure}
Liang, Y., Zhang, Y., Xiong, H., and Sahoo, R. (2007).
\newblock Failure prediction in ibm bluegene/l event logs.
\newblock In {\em Seventh IEEE International Conference on Data Mining (ICDM
  2007)\/}, pages 583--588. IEEE.

\bibitem[Liao {\em et~al.}(2020)Liao, Chen, Li, Zeng, Shang, Guo, Sporea, Toma,
  and Sajedi]{liao2020using}
Liao, L., Chen, J., Li, H., Zeng, Y., Shang, W., Guo, J., Sporea, C., Toma, A.,
  and Sajedi, S. (2020).
\newblock Using black-box performance models to detect performance regressions
  under varying workloads: an empirical study.
\newblock {\em Empirical Software Engineering\/}, {\bf 25}(5), 4130--4160.

\bibitem[Liu {\em et~al.}(2012)Liu, Ting, and Zhou]{liu2012isolation}
Liu, F.~T., Ting, K.~M., and Zhou, Z.-H. (2012).
\newblock Isolation-based anomaly detection.
\newblock {\em ACM Transactions on Knowledge Discovery from Data (TKDD)\/},
  {\bf 6}(1), 1--39.

\bibitem[Liu {\em et~al.}(2022)Liu, Zhang, He, Zhang, Li, Kang, Xu, Ma, Lin,
  Dang, {\em et~al.}]{liu2022uniparser}
Liu, Y., Zhang, X., He, S., Zhang, H., Li, L., Kang, Y., Xu, Y., Ma, M., Lin,
  Q., Dang, Y., {\em et~al.} (2022).
\newblock Uniparser: A unified log parser for heterogeneous log data.
\newblock In {\em Proceedings of the ACM Web Conference 2022\/}, pages
  1893--1901.

\bibitem[Lou {\em et~al.}(2010)Lou, Fu, Yang, Xu, and Li]{lou2010mining}
Lou, J.-G., Fu, Q., Yang, S., Xu, Y., and Li, J. (2010).
\newblock Mining invariants from console logs for system problem detection.
\newblock In {\em 2010 USENIX Annual Technical Conference (USENIX ATC 10)\/}.

\bibitem[Lu {\em et~al.}(2018)Lu, Wei, Li, and Wang]{lu2018detecting}
Lu, S., Wei, X., Li, Y., and Wang, L. (2018).
\newblock Detecting anomaly in big data system logs using convolutional neural
  network.
\newblock In {\em 2018 IEEE 16th Intl Conf on Dependable, Autonomic and Secure
  Computing, 16th Intl Conf on Pervasive Intelligence and Computing, 4th Intl
  Conf on Big Data Intelligence and Computing and Cyber Science and Technology
  Congress (DASC/PiCom/DataCom/CyberSciTech)\/}, pages 151--158. IEEE.

\bibitem[Lyu {\em et~al.}(2021)Lyu, Li, Sayagh, Jiang, and
  Hassan]{lyu2021empirical}
Lyu, Y., Li, H., Sayagh, M., Jiang, Z.~M., and Hassan, A.~E. (2021).
\newblock An empirical study of the impact of data splitting decisions on the
  performance of aiops solutions.
\newblock {\em ACM Transactions on Software Engineering and Methodology
  (TOSEM)\/}, {\bf 30}(4), 1--38.

\bibitem[Meng {\em et~al.}(2019)Meng, Liu, Zhu, Zhang, Pei, Liu, Chen, Zhang,
  Tao, Sun, and Zhou]{meng2019loganomaly}
Meng, W., Liu, Y., Zhu, Y., Zhang, S., Pei, D., Liu, Y., Chen, Y., Zhang, R.,
  Tao, S., Sun, P., and Zhou, R. (2019).
\newblock Loganomaly: Unsupervised detection of sequential and quantitative
  anomalies in unstructured logs.
\newblock In {\em Proceedings of the Twenty-Eighth International Joint
  Conference on Artificial Intelligence, {IJCAI-19}\/}, pages 4739--4745.
  International Joint Conferences on Artificial Intelligence Organization.

\bibitem[Meng {\em et~al.}(2021)Meng, Liu, Zhang, Zaiter, Zhang, Huang, Yu,
  Zhang, Song, Zhang, {\em et~al.}]{meng2021logclass}
Meng, W., Liu, Y., Zhang, S., Zaiter, F., Zhang, Y., Huang, Y., Yu, Z., Zhang,
  Y., Song, L., Zhang, M., {\em et~al.} (2021).
\newblock Logclass: Anomalous log identification and classification with
  partial labels.
\newblock {\em IEEE Transactions on Network and Service Management\/}, {\bf
  18}(2), 1870--1884.

\bibitem[Nagaraj {\em et~al.}(2012)Nagaraj, Killian, and
  Neville]{nagaraj2012structured}
Nagaraj, K., Killian, C., and Neville, J. (2012).
\newblock Structured comparative analysis of systems logs to diagnose
  performance problems.
\newblock In {\em 9th USENIX Symposium on Networked Systems Design and
  Implementation (NSDI 12)\/}, pages 353--366.

\bibitem[Nedelkoski {\em et~al.}(2020)Nedelkoski, Bogatinovski, Acker, Cardoso,
  and Kao]{nedelkoski2020self}
Nedelkoski, S., Bogatinovski, J., Acker, A., Cardoso, J., and Kao, O. (2020).
\newblock Self-attentive classification-based anomaly detection in unstructured
  logs.
\newblock In {\em 2020 IEEE International Conference on Data Mining (ICDM)\/},
  pages 1196--1201. IEEE.

\bibitem[Nguyen {\em et~al.}(2016)Nguyen, Walde, and Vu]{nguyen2016integrating}
Nguyen, K.~A., Walde, S. S.~i., and Vu, N.~T. (2016).
\newblock Integrating distributional lexical contrast into word embeddings for
  antonym-synonym distinction.
\newblock {\em arXiv preprint arXiv:1605.07766\/}.

\bibitem[Oliner and Stearley(2007)Oliner and
  Stearley]{oliner2007supercomputers}
Oliner, A. and Stearley, J. (2007).
\newblock What supercomputers say: A study of five system logs.
\newblock In {\em 37th annual IEEE/IFIP international conference on dependable
  systems and networks (DSN'07)\/}, pages 575--584. IEEE.

\bibitem[Oliner {\em et~al.}(2012)Oliner, Ganapathi, and
  Xu]{oliner2012advances}
Oliner, A., Ganapathi, A., and Xu, W. (2012).
\newblock Advances and challenges in log analysis.
\newblock {\em Communications of the ACM\/}, {\bf 55}(2), 55--61.

\bibitem[Prewett(2003)Prewett]{prewett2003analyzing}
Prewett, J.~E. (2003).
\newblock Analyzing cluster log files using logsurfer.
\newblock In {\em Proceedings of the 4th Annual Conference on Linux
  Clusters\/}. Citeseer.

\bibitem[Rouillard(2004)Rouillard]{rouillard2004real}
Rouillard, J.~P. (2004).
\newblock Real-time log file analysis using the simple event correlator (sec).
\newblock In {\em LISA\/}, volume~4, pages 133--150.

\bibitem[Rusticus and Lovato(2014)Rusticus and Lovato]{rusticus2014impact}
Rusticus, S.~A. and Lovato, C.~Y. (2014).
\newblock Impact of sample size and variability on the power and type i error
  rates of equivalence tests: A simulation study.
\newblock {\em Practical Assessment, Research, and Evaluation\/}, {\bf 19}(1),
  11.

\bibitem[Salton and Buckley(1988)Salton and Buckley]{salton1988term}
Salton, G. and Buckley, C. (1988).
\newblock Term-weighting approaches in automatic text retrieval.
\newblock {\em Information processing \& management\/}, {\bf 24}(5), 513--523.

\bibitem[Schroeder and Gibson(2007)Schroeder and
  Gibson]{Schroeder07DiskFailures}
Schroeder, B. and Gibson, G.~A. (2007).
\newblock Disk failures in the real world: What does an {MTTF} of 1,000,000
  hours mean to you?
\newblock In {\em 5th USENIX Conference on File and Storage Technologies (FAST
  07)\/}, San Jose, CA. USENIX Association.

\bibitem[Shang {\em et~al.}(2014)Shang, Jiang, Adams, Hassan, Godfrey, Nasser,
  and Flora]{shang2014exploratory}
Shang, W., Jiang, Z.~M., Adams, B., Hassan, A.~E., Godfrey, M.~W., Nasser, M.,
  and Flora, P. (2014).
\newblock An exploratory study of the evolution of communicated information
  about the execution of large software systems.
\newblock {\em Journal of Software: Evolution and Process\/}, {\bf 26}(1),
  3--26.

\bibitem[Tantithamthavorn {\em et~al.}(2017)Tantithamthavorn, McIntosh, Hassan,
  and Matsumoto]{tantithamthavorn2017mvt}
Tantithamthavorn, C., McIntosh, S., Hassan, A.~E., and Matsumoto, K. (2017).
\newblock An empirical comparison of model validation techniques for defect
  prediction models.
\newblock {\em IEEE Transactions on Software Engineering\/}, {\bf 43}(1),
  1--18.

\bibitem[Tantithamthavorn {\em et~al.}(2018)Tantithamthavorn, McIntosh, Hassan,
  and Matsumoto]{tantithamthavorn2018impact}
Tantithamthavorn, C., McIntosh, S., Hassan, A.~E., and Matsumoto, K. (2018).
\newblock The impact of automated parameter optimization on defect prediction
  models.
\newblock {\em IEEE Transactions on Software Engineering\/}, {\bf 45}(7),
  683--711.

\bibitem[Turc {\em et~al.}(2019)Turc, Chang, Lee, and Toutanova]{turc2019}
Turc, I., Chang, M.-W., Lee, K., and Toutanova, K. (2019).
\newblock Well-read students learn better: On the importance of pre-training
  compact models.
\newblock {\em arXiv preprint arXiv:1908.08962v2\/}.

\bibitem[van~der Maaten and Hinton(2008)van~der Maaten and
  Hinton]{van2008visualizing}
van~der Maaten, L. and Hinton, G. (2008).
\newblock Visualizing data using t-sne.
\newblock {\em Journal of Machine Learning Research\/}, {\bf 9}(86),
  2579--2605.

\bibitem[Vaswani {\em et~al.}(2017)Vaswani, Shazeer, Parmar, Uszkoreit, Jones,
  Gomez, Kaiser, and Polosukhin]{vaswani2017attention}
Vaswani, A., Shazeer, N., Parmar, N., Uszkoreit, J., Jones, L., Gomez, A.~N.,
  Kaiser, L.~u., and Polosukhin, I. (2017).
\newblock Attention is all you need.
\newblock In I.~Guyon, U.~V. Luxburg, S.~Bengio, H.~Wallach, R.~Fergus,
  S.~Vishwanathan, and R.~Garnett, editors, {\em Advances in Neural Information
  Processing Systems\/}, volume~30. Curran Associates, Inc.

\bibitem[Wan {\em et~al.}(2021)Wan, Liu, Wang, and Wen]{wan2021glad}
Wan, Y., Liu, Y., Wang, D., and Wen, Y. (2021).
\newblock Glad-paw: Graph-based log anomaly detection by position aware
  weighted graph attention network.
\newblock In {\em Advances in Knowledge Discovery and Data Mining: 25th
  Pacific-Asia Conference, PAKDD 2021, Virtual Event, May 11--14, 2021,
  Proceedings, Part I\/}, pages 66--77. Springer.

\bibitem[Wang {\em et~al.}(2018)Wang, Xu, and Guo]{wang2018anomaly}
Wang, M., Xu, L., and Guo, L. (2018).
\newblock Anomaly detection of system logs based on natural language processing
  and deep learning.
\newblock In {\em 2018 4th International Conference on Frontiers of Signal
  Processing (ICFSP)\/}, pages 140--144. IEEE.

\bibitem[Xie {\em et~al.}(2022)Xie, Zhang, and Babar]{xie2022loggd}
Xie, Y., Zhang, H., and Babar, M.~A. (2022).
\newblock Loggd: Detecting anomalies from system logs by graph neural networks.
\newblock {\em arXiv preprint arXiv:2209.07869\/}.

\bibitem[Xu {\em et~al.}(2009)Xu, Huang, Fox, Patterson, and
  Jordan]{xu2009detecting}
Xu, W., Huang, L., Fox, A., Patterson, D., and Jordan, M.~I. (2009).
\newblock Detecting large-scale system problems by mining console logs.
\newblock In {\em Proceedings of the ACM SIGOPS 22nd symposium on Operating
  systems principles\/}, pages 117--132.

\bibitem[Yuan {\em et~al.}(2010)Yuan, Mai, Xiong, Tan, Zhou, and
  Pasupathy]{yuan2010sherlog}
Yuan, D., Mai, H., Xiong, W., Tan, L., Zhou, Y., and Pasupathy, S. (2010).
\newblock Sherlog: error diagnosis by connecting clues from run-time logs.
\newblock In {\em Proceedings of the fifteenth International Conference on
  Architectural support for programming languages and operating systems\/},
  pages 143--154.

\bibitem[Yuan {\em et~al.}(2012)Yuan, Park, and Zhou]{yuan2012characterizing}
Yuan, D., Park, S., and Zhou, Y. (2012).
\newblock Characterizing logging practices in open-source software.
\newblock In {\em 2012 34th International Conference on Software Engineering
  (ICSE)\/}, pages 102--112. IEEE.

\bibitem[Zhang {\em et~al.}(2019)Zhang, Xu, Lin, Qiao, Zhang, Dang, Xie, Yang,
  Cheng, Li, {\em et~al.}]{zhang2019robust}
Zhang, X., Xu, Y., Lin, Q., Qiao, B., Zhang, H., Dang, Y., Xie, C., Yang, X.,
  Cheng, Q., Li, Z., {\em et~al.} (2019).
\newblock Robust log-based anomaly detection on unstable log data.
\newblock In {\em Proceedings of the 2019 27th ACM Joint Meeting on European
  Software Engineering Conference and Symposium on the Foundations of Software
  Engineering\/}, pages 807--817.

\bibitem[Zhu {\em et~al.}(2019)Zhu, He, Liu, He, Xie, Zheng, and
  Lyu]{zhu2019tools}
Zhu, J., He, S., Liu, J., He, P., Xie, Q., Zheng, Z., and Lyu, M.~R. (2019).
\newblock Tools and benchmarks for automated log parsing.
\newblock In {\em 2019 IEEE/ACM 41st International Conference on Software
  Engineering: Software Engineering in Practice (ICSE-SEIP)\/}, pages 121--130.
  IEEE.

\end{thebibliography}

\end{document}